\documentclass[twocolumn,aps,prb,superscriptaddress]{revtex4-2}
\usepackage{graphicx}
\usepackage{bm}
\usepackage[linktocpage=true,colorlinks=true,urlcolor=blue,linkcolor=red,citecolor = blue]{hyperref}
\usepackage{amsfonts}
\usepackage{amsmath}
\usepackage{amssymb}
\usepackage{braket}
\usepackage{newfloat}
\usepackage{siunitx}
\usepackage{physics}
\usepackage{mathtools}
\usepackage{xcolor}
\usepackage{comment}
\usepackage[normalem]{ulem}
\newcommand{\red}[1]{\textcolor{black}{#1}}
\newcommand{\T}[1]{\text{#1}}
\usepackage[T1]{fontenc}

\usepackage{titlesec}
\titleformat{\section}{\normalfont\large\bfseries}{\thesection}{1em}{}

\begin{document}

\title{Reducing strain fluctuations in quantum dot devices by gate-layer stacking}

\author{Collin C. D. Frink}
\altaffiliation[These authors ]{contributed equally.}
\affiliation{Department of Physics, University of Wisconsin-Madison, Madison, Wisconsin 53706, USA}
\author{Talise Oh}
\altaffiliation[These authors ]{contributed equally.}
\affiliation{Department of Physics, University of Wisconsin-Madison, Madison, Wisconsin 53706, USA}
\author{E. S. Joseph}
\affiliation{Department of Physics, University of Wisconsin-Madison, Madison, Wisconsin 53706, USA}
\author{Merritt P. Losert}
\affiliation{Department of Physics, University of Wisconsin-Madison, Madison, Wisconsin 53706, USA}
\author{E. R. MacQuarrie}
\affiliation{Department of Physics, University of Wisconsin-Madison, Madison, Wisconsin 53706, USA}
\author{Benjamin D. Woods}
\email[Corresponding author: ] {bwoods6@wisc.edu}
\affiliation{Department of Physics, University of Wisconsin-Madison, Madison, Wisconsin 53706, USA}
\author{M. A. Eriksson}
\affiliation{Department of Physics, University of Wisconsin-Madison, Madison, Wisconsin 53706, USA}
\author{Mark Friesen}
\affiliation{Department of Physics, University of Wisconsin-Madison, Madison, Wisconsin 53706, USA}

\maketitle

\noindent \textbf{Abstract.} \noindent
Nanofabricated metal gate electrodes are commonly used to confine and control electrons in electrostatically defined quantum dots. 
However, these same gates impart strain-induced potential fluctuations that can potentially impair device functionality.
Here we investigate strain fluctuations in Si/SiGe heterostructures, caused by (i) lattice mismatch, (ii) materials-dependent thermal contraction, and (iii) depositional stress in the metal gates. 
By simulating gate geometries, ranging from simple to realistically complicated, we identify two opposing effects in overlapping gate structures: (a) gate-driven behavior arising from isolated gates vs (b) oxide-driven behavior arising from the thin oxides separating the gates in an overlapping geometry.
These limiting behaviors induce strains of opposite sign, pointing towards the possibility of suppressing strain fluctuations through  careful design.
Here, we demonstrate \red{in numerical models, a} nearly total suppression of short-range strain fluctuation through device optimization.
These results suggest that strain fluctuations should not pose an insurmountable challenge to qubit uniformity, provided that oxide and overlapping gate thicknesses can be tuned.

\vspace{.5in}

Gate-defined quantum dots in Si/SiGe quantum wells are a promising platform for large-scale quantum computation~\cite{Loss1998,Zwanenburg2013,Burkard2023}, where recent demonstrations of single and two qubit gates have exceeded error correction thresholds~\cite{Xue2022,Noiri2022,Mills2021}. 
While these achievements represent important milestones, useful  quantum hardware will require vast arrays of reliable, low-error qubits.
Scaling up will require a level of qubit reproducibility and uniformity on par with transistors in modern integrated circuits~\cite{Zwerver2022,Meyer2023}.
Such uniformity has been achieved for certain dot properties like orbital and charging energies~\cite{Zajac2016}, but remains a challenge for properties like the valley energy splitting~\cite{PRApplied044033}, due to the inherent atomistic disorder of the SiGe random alloy~\cite{PaqueletWuetz2022,Losert2023}.
Some types of non-uniformity are potentially reconfigurable, such as interfacial trapped charge~\cite{meyer2023b, Wolfe:2024p034044}, which can modify the local electrostatics~\cite{Massai2023} and cause the formation of unintentional quantum dots~\cite{PhysRevB.80.075310,PhysRevB.80.115331,Hu2009}.
Other sources of variability (e.g., valley splitting) are immutable, and are prescribed during heterostructure growth or device fabrication.

\begin{figure}
\begin{center}
\includegraphics[width=0.47\textwidth]{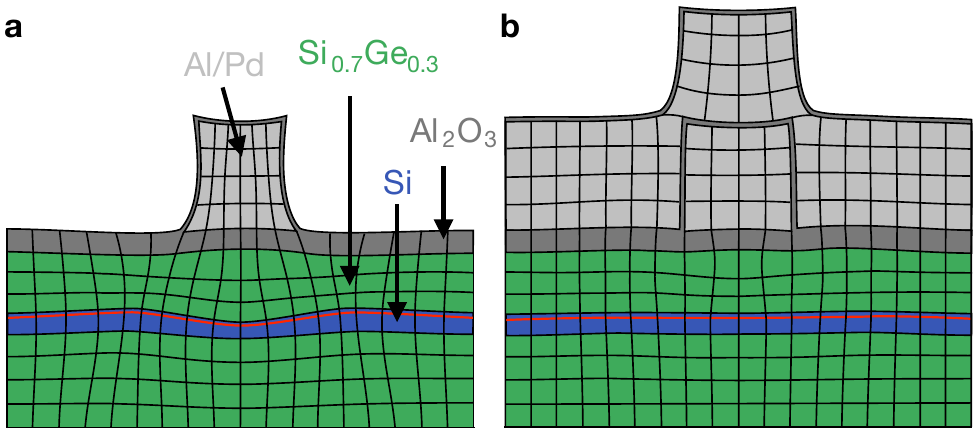}
\end{center}
\vspace{-0.5cm}
\hspace{-1cm}
\caption{\textbf{Schematic illustration of the suppression of strain fluctuations by gate-layer stacking.} 
\textbf{a} Non-uniform strain in a Si/SiGe quantum well (blue/green) is induced beneath a metal wire (light gray) upon cooling the device, due to differing thermal expansion coefficients in the materials. 
The red line indicates the plane of the two-dimensional electron gas (2DEG).
\textbf{b} Same as \textbf{a}, with a global gate covering the original wire. If the oxide layer (dark gray) between the gates is thin enough, and the overlapping gate is thick enough, the combined structure imparts a nearly uniform strain field to the quantum well below.
For thinner structures, oxide and gate thicknesses can be carefully tuned to yield similar uniformity. Note that strains and structural deformations have been exaggerated in these diagrams, for visual clarity; see simulations, below, for numerically accurate results.} 
\label{FIG_Cartoon}
\vspace{-1mm}
\end{figure}

Local strain fields can significantly affect the in-plane confinement potential of quantum dots~\cite{Thorbeck,Park2016}.
These strains arise from the structural or geometrical features of a device, such as metal electrodes or etched regions, which are intentionally patterned atop the quantum well that houses the qubits~\footnote{Strain fluctuations also arise from misfit locations in plastically relaxed virtual substrates~\cite{CorleyWiciak2023}; however, we do not consider such effects here.}.
Such structures are carefully designed to provide electrostatic control of the qubit environment.
However, the strains arising from these gates can have a very different effect on the confinement potential than the intended one, resulting in energy variations on the order of meV \cite{Stein2021, Mooy2020}, which is comparable to the electrostatic confinement. 
These fluctuations can shift the positions of quantum dots~\cite{Pateras2018}, cause the formation of unintended dots~\cite{Thorbeck}, affect their exchange interactions~\cite{Burkard2023,Hu2006,Deng2018,Deng2020}, or modify their magnetic properties~\cite{Liles2021,Uriel2023}.

\begin{figure*}[t]
\begin{center}
\includegraphics[width=\textwidth]{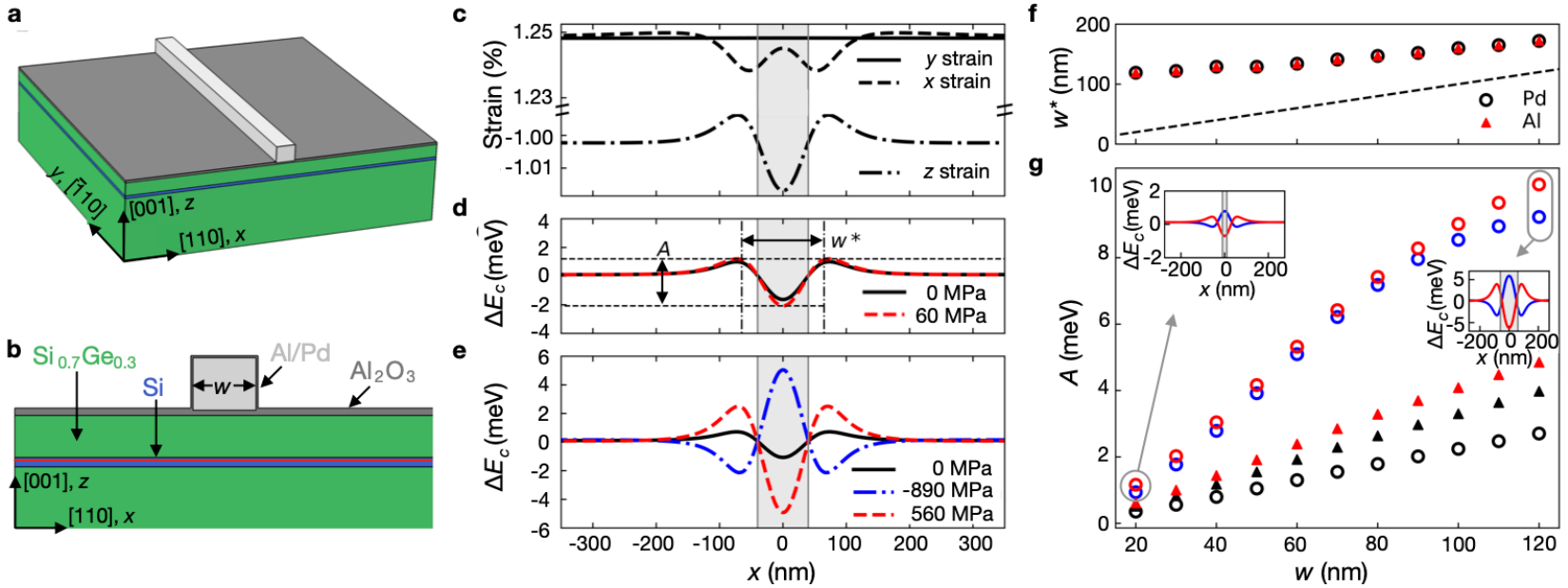}
\end{center}
\vspace{-1cm}
\hspace{-1cm}
\caption{
\textbf{Strain simulations of a single-wire geometry.} 
\textbf{a,b} 3D and cross-sectional views, respectively, with the coordinate axes indicated.
From bottom to top, the thicknesses of the Si$_{0.7}$Ge$_{0.3}$ virtual substrate (green), Si quantum well (blue), Si$_{0.7}$Ge$_{0.3}$ spacer (green), and insulating Al$_2$O$_3$ layer (dark gray) are 2~$\mu$m, 9~nm, 40~nm, and 10~nm, respectively. Three additional heterostructure configurations are considered in supplementary Fig.~S7.
A metal wire (light gray) of height \red{70~nm} and variable width $w$ is formed of Al or Pd.
The wire is covered on the top and sides by a thin 2~nm Al$_2$O$_3$ layer (dark gray; not shown in \textbf{a}, for clarity). 
All strain and energy fluctuations in this work are evaluated on the horizontal red line in \textbf{b}, corresponding to the plane of the 2DEG, which is taken to lie 1.5~nm below the top quantum-well interface.
\textbf{c} The diagonal components of the strain tensor, $\varepsilon_{xx}$, $\varepsilon_{yy}$, and $\varepsilon_{zz}$ are shown for the case of a single Al wire of width $w = 80$~nm.
\textbf{d,e} Corresponding conduction-band energy offsets $\Delta E_c$, for Al (\textbf{d}) or Pd (\textbf{e}) wires of width $w=80$~nm (shaded region), with depositional stresses as indicated.
We also assume thermal contractions appropriate for a temperature of 1~K, as described in Methods. 
The peak-to-peak width $w^*$ and amplitude $A$ of the fluctuations are defined in \textbf{d}. 
\textbf{f,g} Results for $w^*$ and $A$ are plotted as a function of the actual wire width $w$.
The dashed line in \textbf{f} corresponds to $w^* = w$.
Closed triangles in \textbf{g} correspond to Al wires, following the color scheme in \textbf{d}, while open circles correspond to Pd, following the color scheme in \textbf{e}. 
Insets show energy shifts orthogonal to the wire, for the indicated Pd      wire parameters.
}
\label{FIG_SingleWire}
\vspace{-1mm}
\end{figure*}

In this work, we numerically investigate strain effects arising from three main sources: (i) lattice mismatch between the Si in the quantum well and the SiGe alloy of the barriers~\cite{Schaffler1997}, (ii) unequal contractions of different materials as the device is cooled~\cite{Thorbeck}, and (iii) depositional stress, which occurs for example when metal is deposited on a semiconductor or an oxide~\cite{Abadias2018}. For numerical accuracy, we also adopt a more-realistic elasticity model than the simple isotropic models often assumed for Si-based devices. 
Specifically, we consider an orthotropic elastic tensor, as appropriate for [001]-oriented Si/SiGe heterostructures~\cite{Hopcroft}. 
Comparing our results to the standard, isotropic treatments, we observe relative variations in the gate-induced strain-fields greater than \red{24\%}, for typical quantum dot devices. 
(See supplementary Figs.~S1 and S2.)
These results indicate that more-accurate simulations should always incorporate the correct orthotropic strain model.

We employ such strain calculations to estimate conduction-band energies in the Si quantum well, using the conversion described in Methods, from which we deduce the locally varying potential landscape.
(In supplementary Fig.~S5, we also include the corresponding electrostatic potential.)
Beginning with simple gate structures, we work our way up to realistic devices measured in qubit experiments~\cite{Neyens2019,Xue2021}.
Due to the long-range nature of the strain fields, our simulations exhibit both short-range variations that mirror the locally varying gate structure, as well as smoother, averaged behavior arising from multiple gates, which is manifested (for example) as edge effects.

By exploring gate designs commonly employed in experiments, we can draw some practical conclusions for designing future experiments. Most notably, we find that undesirable strain fluctuations can be greatly suppressed by stacking multiple layers of gates while carefully tuning the thicknesses of the gate and oxide layers.
To illustrate this point, in Fig.~\ref{FIG_Cartoon}a, we depict the strong, locally varying strain found in the plane of the quantum well due to an isolated, narrow gate. 
In contrast, it is clear that a wide gate will produce a relatively uniform strain pattern in the quantum well below, except near its edges.
Now if we consider an overlapping geometry like the one shown in Fig.~\ref{FIG_Cartoon}b, where the top gate is very thick and the oxide is very thin, the strain in the quantum well will be approximately uniform, similar to the situation for a single, wide gate. 
Importantly, the residual non-uniformity of the strain in Fig.~\ref{FIG_Cartoon}b arises from the oxide layer, while the non-uniformity in Fig.~\ref{FIG_Cartoon}a is from the isolated gate. 
In this work, we designate these two extreme behaviors as gate-driven (as in Fig.~\ref{FIG_Cartoon}a) or oxide-driven (as in Fig.~\ref{FIG_Cartoon}b), where the latter causes strain of the opposite sign.
We further show that short-range strain fluctuations can be strongly suppressed by carefully tuning the layer thicknesses to the crossover between these two regimes.
In the following discussion, we elaborate on these and other phenomena and provide additional numerical details.

\begin{figure}[t]
\begin{center}
\includegraphics[width=0.48\textwidth]{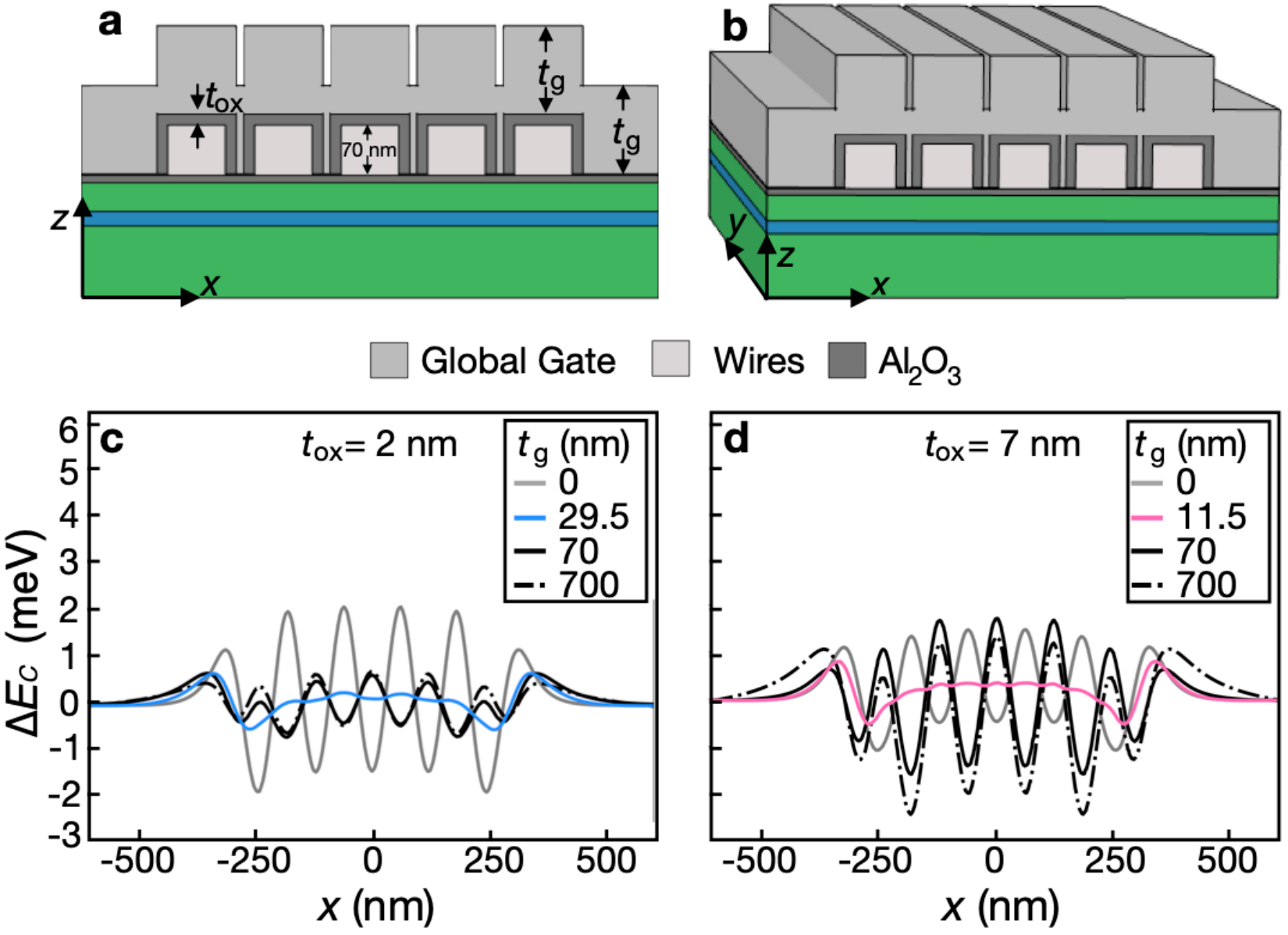}
\end{center}
\vspace{-1cm}
\hspace{-1.0cm}
\caption{
\textbf{Strain simulations of a parallel-wire geometry, with and without a large ($18\times 18$~$\mu$m$^2$) global gate of variable thickness $t_g$, fabricated on the same heterostructure as Fig.~\ref{FIG_SingleWire}.}
(Figures are not to scale.) Strain fields are shown in  supplementary Fig.~S9.
\textbf{a} A cross-sectional view of five parallel Al wires (80~nm wide, 70~nm tall, with 40~nm gaps between them). 
The wires are separated from the global gate by a layer of Al$_2$O$_3$ with variable thickness, $t_\text{ox}$.
\textbf{b} A 3D depiction of the same geometry. 
In both \textbf{a} and \textbf{b}, the topmost 2~nm oxide layer is not shown, for clarity.
\textbf{c} Conduction-band energy modulations $\Delta E_c$, for the geometry depicted in \textbf{a,b}, with varying $t_g$ values and $t_\text{ox}=2$~nm. 
\textbf{d} Same as \textbf{c}, but with a $t_\text{ox}=7$~nm.
\textbf{c,d} Dashed black lines show results for a global gate of thickness $t_g=700$~nm, which is ten times greater than the height of the five wires. 
The results are very similar to those from a much thinner global gate (solid black lines), with the same thickness as the five wires,  $t_g=70$~nm.
The gray lines show results for the case of no global gate, $t_g=0$. 
The blue and pink lines show results for the optimized geometries that minimize fluctuations, corresponding to \red{$t_g=29.5$~nm} in \textbf{c} and $t_g=11.5$~nm in \textbf{d}.
}
\label{FIG_MultiWire}
\vspace{-1ex}
\end{figure}

\begin{figure}
\begin{center}
\includegraphics[width=1\linewidth]{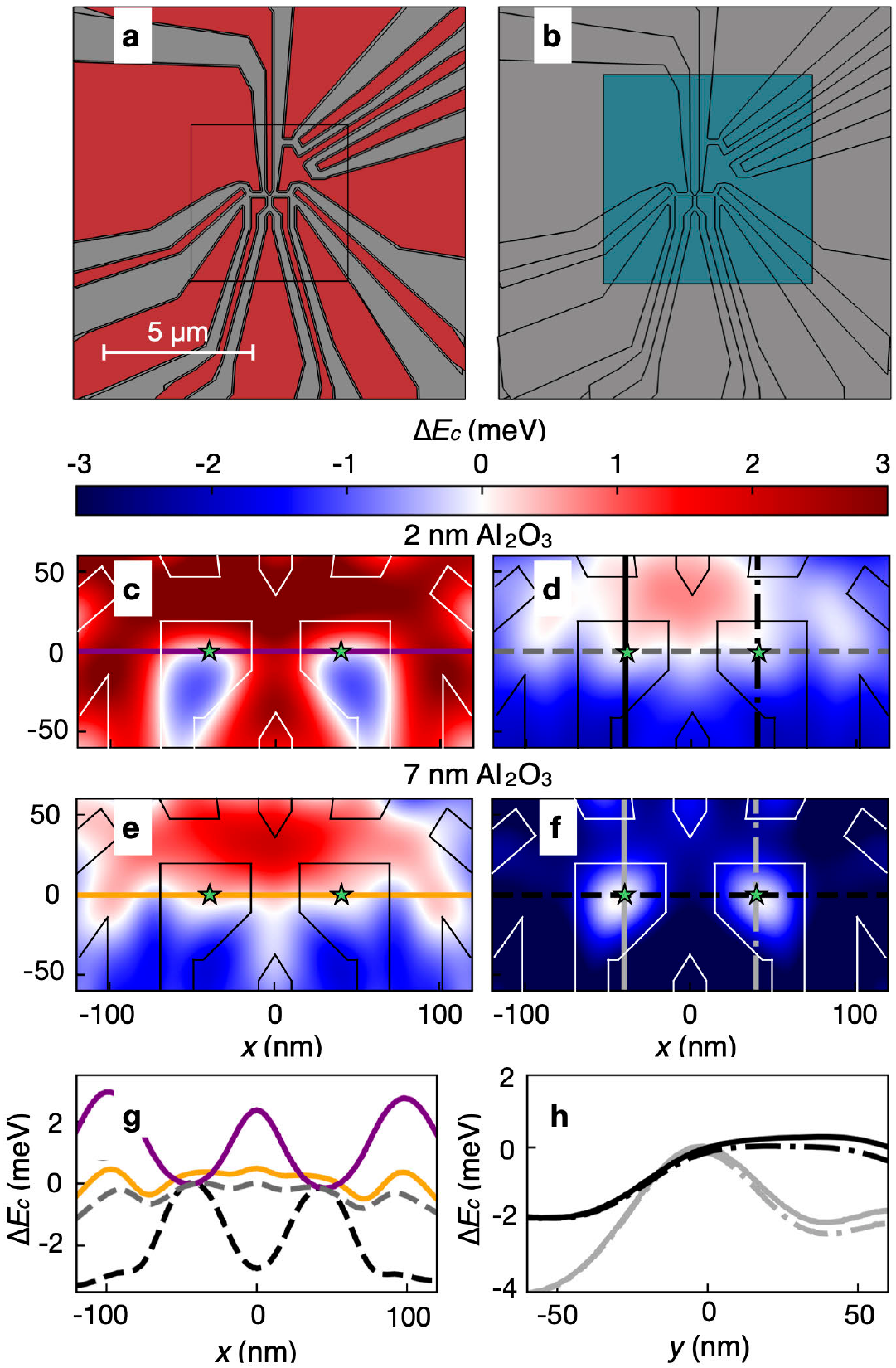}
\end{center}
\vspace{-1cm}
\hspace{-1cm}
\caption{
\textbf{Strain simulations of the two-dot device of~\cite{Xue2021}, for two configurations of stacked Al gates, and two oxide thicknesses.}
From bottom to top, the thicknesses of the Si$_{0.7}$Ge$_{0.3}$ virtual substrate, Si quantum well, Si$_{0.7}$Ge$_{0.3}$ spacer, and Si cap are 1.2~$\mu$m, 8~nm, 30~nm, and 1~nm, respectively.
\textbf{a} Reduced gate set, including only the lower layer of gates (20~nm thick), is shown in red. 
We also include Al$_2$O$_3$ layers of variable thickness, above and below the lower gate layer (top oxide layer is not shown, for clarity).
\textbf{b} Same as \textbf{a}, but including a global top gate (turquoise) of thickness 40~nm.
\textbf{c,d} Strain-induced fluctuations of $\Delta E_c$ in the plane of the 2DEG, for the geometries shown in \textbf{a} and \textbf{b}, respectively, where all oxide layers have a thickness of 2~nm.
\textbf{e,f} Same as \textbf{c} and \textbf{d}, except the oxide layers have a thickness of 7~nm.
In \textbf{c}-\textbf{f}, approximate dot locations are indicated by green stars, and we have shifted the energy scale such that $\Delta E_c=0$ at the dot centers. 
\textbf{g} Horizontal linecuts, indicated in panels \textbf{c}-\textbf{f}, with appropriate color codings.
We again highlight a key result: the linecut most closely associated with gate-driven behavior (solid--purple line) exhibits fluctuations with opposite sign as the linecut most associated with oxide-driven behavior (dashed-black line).
\textbf{h} Vertical linecuts, indicated in panels \textbf{d} and \textbf{f}, with appropriate color codings.}
\label{FIG_SingleLayer}
\vspace{-1mm}
\end{figure}

\begin{figure}
\begin{center}
\includegraphics[width=1\linewidth]{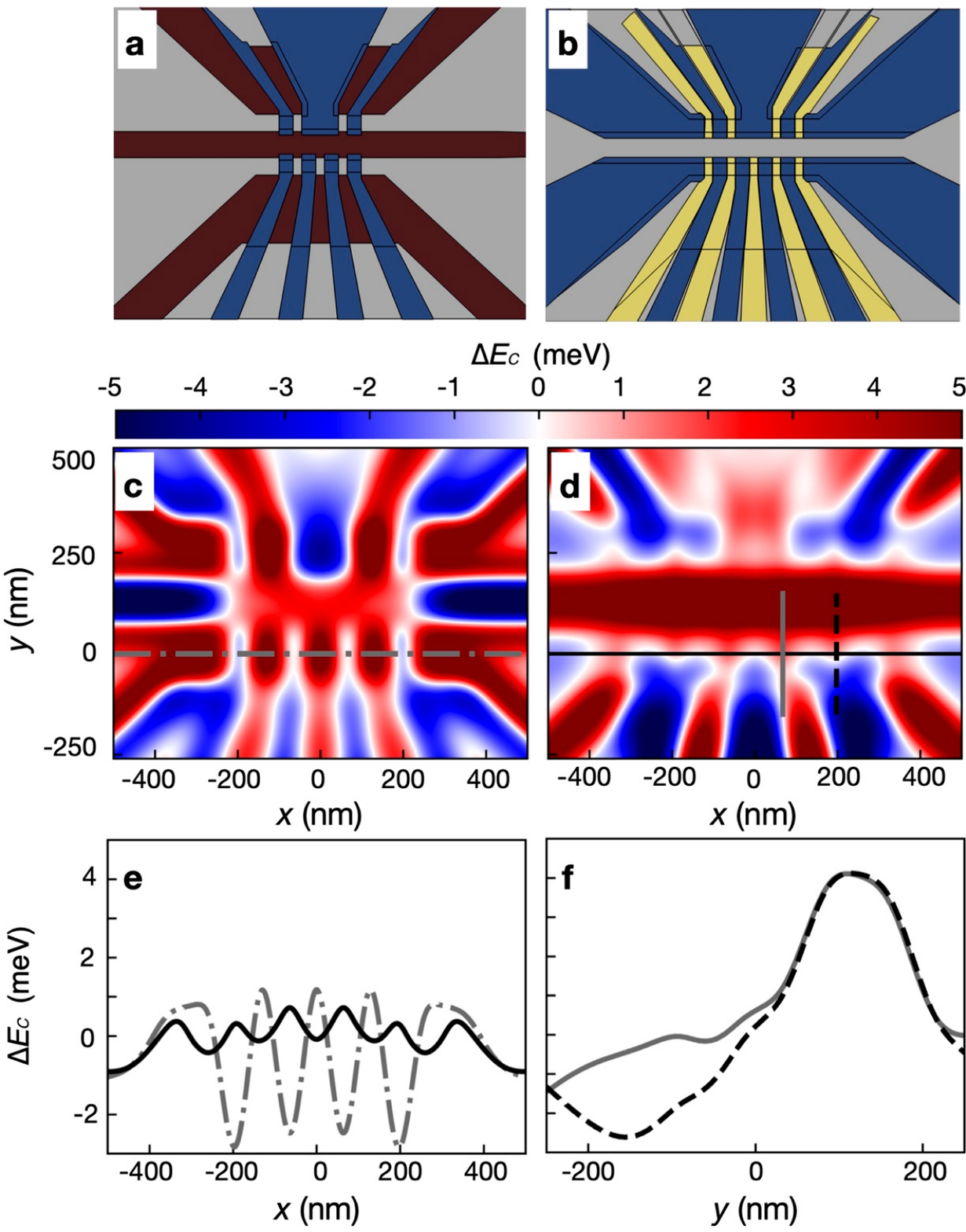}
\end{center}
\vspace{-0.5cm}
\hspace{-1cm}
\caption{
\textbf{Strain simulations of the quadruple-dot device of Ref.~\cite{Neyens2019}, for several configurations of stacked Al gates fabricated on the same heterostructure as Fig.~\ref{FIG_SingleWire}.}
\textbf{a} A reduced gate set, including only screening gates (maroon, $30~\text{nm}$ thick), plunger gates (blue, $50~\text{nm}$ thick), and upper reservoir gates (also shown in blue because they are formed in the same layer as the plungers). 
\textbf{b} The full gate geometry, including the gate layers in \textbf{a}, as well as side-reservoir gates (also blue), and tunnel-barrier gates (yellow, $70~\text{nm}$ thick). 
Two additional gate configurations are considered in supplementary Fig.~S4.
Note that every gate layer is covered by 2~nm of Al$_2$O$_3$, although the topmost layer is not pictured, for clarity.
\textbf{c, d} Strain-induced fluctuations of $\Delta E_c$ in the plane of the 2DEG, for the geometries shown in \textbf{a} and \textbf{b}, respectively. 
\textbf{e} Results for the horizontal linecuts indicated in \textbf{c} (dot-dashed gray line) and \textbf{d} (solid black line), which pass through the centers of the lower four dots. 
\textbf{f} Results for the vertical linecuts indicated in \textbf{d}.}
\label{FIG_Quad}
\end{figure}

\vspace{.2in}
\noindent \textbf{\large Results} \\
To study the physics of strain fluctuations under top gates, and particularly, the effect of gate-layer stacking, we simulate four systems of varying complexity. 
The first system consists of a long metal gate fabricated atop an otherwise uniform Si/SiGe heterostructure, as shown in Fig.~\ref{FIG_SingleWire}a, which serves as a minimal model for describing the effects of local strain. The second system includes several closely spaced gates, with or without an overlaid global gate layer, as illustrated in Fig.~\ref{FIG_MultiWire}.
By varying the model parameters of this simple geometry, we are able to systematically explore the physics of gate-layer stacking.
Figures~\ref{FIG_SingleLayer} and \ref{FIG_Quad} correspond to the experimental devices studied in Refs.~\cite{Xue2021} and \cite{Neyens2019}, respectively. 
Here, the effects of gate-layer stacking are characterized by including different subsets of overlapping gates in the simulations. 
Figure~\ref{FIG_SingleLayer} also highlights the importance of oxide-layer thickness. 
All results shown in this work make use of the Solid Mechanics module of COMSOL Multiphysics~\cite{Comsol}, as described in Methods. 
The simulations incorporate all three sources of strain described above, with the materials parameters described in Methods, and the orthotropic strain tensor described in supplementary Sec.~S1.
For the thermal contraction simulations, we assume the devices are cooled from $T = 293.15$ to $1~\text{K}$. 

\vspace{.2in}
\noindent \textbf{Single wire} \\
We first consider the simple wire geometry shown in Fig.~\ref{FIG_SingleWire}a, where the wire is uniform along the $\hat y$ direction.(In this work, we adopt the coordinate system $\hat x=[110]$, \red{$\hat y=[\overline110]$}, and $\hat z=[001]$, as indicated in the figure.)
The corresponding device cross-section is shown in Fig.~\ref{FIG_SingleWire}b.
From bottom to top, the heterostructure in our simulations consists of a 2~$\mu$m layer of strain-relaxed Si\textsubscript{$0.7$}Ge\textsubscript{$0.3$}, a 9~nm Si quantum well, another 40~nm spacer layer of Si\textsubscript{$0.7$}Ge\textsubscript{$0.3$}, and a 10~nm insulating layer of Al$_2$O$_3$.
We assume that all Si or SiGe interfaces in this structure are grown epitaxially, and that the quantum well is fully strained. 
The Si layer therefore experiences biaxial tensile strain due to the larger bulk lattice constant of SiGe.
Except where noted (Fig.~\ref{FIG_SingleLayer}), we use this same heterostructure in all simulations reported below.
Moreover the strain, and the energy variations it causes, are always evaluated in the plane of the two-dimensional electron gas (2DEG), which we take to lie 1.5~nm below the top quantum-well interface. 

We first simulate just the heterostructure geometry, without any top gate.
Here, the strain is caused by a combination of lattice mismatch and thermal contraction, giving the following diagonal strain results in the plane of the 2DEG: $\varepsilon_{xx} = \varepsilon_{yy} = 1.2707~\%$ and $\varepsilon_{zz} = -0.9798~\%$.
This strain causes an energy shift of the conduction band, as described in Methods, which accounts (in part) for the vertical confinement potential of the quantum well.
We take as a reference point the conduction-band minimum at a point far away from the region of interest.
Since the strain and the conduction-band minimum are both uniform in this geometry, the energy shift compared to the reference point, $\Delta E_c$, is zero across the whole sample.

For the wire geometry of Figs.~\ref{FIG_SingleWire}a, b, we consider an Al metal wire formed directly atop the oxide layer as shown, with a fixed height of \red{70~nm} and a variable width $w$.
The wire causes local strain fluctuations, due to a combination of depositional stress and thermal contractions, with results shown in Fig.~\ref{FIG_SingleWire}c. 
Here all three diagonal strain-tensor components are plotted ($\varepsilon_{xx}$, $\varepsilon_{yy}$, and $\varepsilon_{zz}$).
The corresponding conduction-band energy shifts are shown in Fig.~\ref{FIG_SingleWire}d, for cases with and without depositional stress, as indicated, where we assume a depositional stress of $60~\T{MPa}$~\cite{Guisbiers2006}.
Similarly, Fig.~\ref{FIG_SingleWire}e shows results for a Pd wire.
In this case, different values (and signs) of the depositional stress are reported in the literature\cite{Guisbiers2006,Afshar2010};
we therefore consider two different values here, as well the case of no depositional stress, as indicated in the figure.
(The results with no depositional stress differ from those in the Al wire because of the different thermal contractions.)
For both sets of simulations, the energy shift converges to its asymptotic value far from the wire, where we define $\Delta E_c=0$.
Near the gates, $\Delta E_c$ has strong variations on the scale of several meV. 
For the Al wire, the depositional stress is low, and we see that its contribution to $\Delta E_c$ is small. 
For the Pd wire, the opposite is true, and the depositional stress dominates over the thermal stress.
Here, $\Delta E_c$ changes sign for the case of depositional tensile stress ($560~\text{MPa}$) vs compressive stress ($-890~\text{MPa}$). 
supplementary Fig.~S3 provides further insight into the specific shape of the strain profiles.  

The dependence of strain effects on wire width is shown in Figs.~\ref{FIG_SingleWire}f, g.
We now introduce two parameters to characterize the $\Delta E_c$ variations, as defined in Fig.~\ref{FIG_SingleWire}d: the peak-to-peak width $w^*$ and the trough-to-peak amplitude $A$.
These two parameters are plotted in Figs.~\ref{FIG_SingleWire}f, g, respectively. The width parameter $w^*$ is found to be essentially universal, with no significant dependence on materials or strain parameters.
Interestingly, $w^*$ asymptotes to a nonzero value for small $w$, and to $w + w_0$ (with $w_0>0$), for large $w$. 
Both effects can be understood in terms of an approximate ``45$^\circ$ rule,'' in which the strain fields extend out at an angle from the gates, with details depending slightly on the wire width.
We note that there is no contradiction that $w^*$ remains nonzero as $w\rightarrow 0$, since $A\rightarrow 0$ also in this limit.
However, there is an interesting dependence of $A$ on the strain parameters and on $w$.
We see that $A$ initially follows a linear dependence on $w$ in all cases. 
For the case of the red and blue circle data, this behavior changes at around $w\approx 80$-90~nm, where $A$ begins to plateau as the two edges of the wire no longer affect each other's local strain field. 
Such plateaus are generally expected for large $w$.

\vspace{.2in}
\noindent \textbf{Parallel wires} \\
We now study how the single-wire picture is modified in the presence of additional wires, spaced closely enough that their strain fields overlap. 
Specifically, we consider the geometry shown in Figs.~\ref{FIG_MultiWire}a and b, comprising five parallel Al wires of width $80~\T{nm}$ and height $70~\T{nm}$, with gaps of $40~\T{nm}$ between the wires.
\red{These choices reflect typical gate dimensions for quantum-dot qubits~\cite{Neyens2019} and are representative of the realistic devices considered in later sections of this work.}
Through a series of simulations we study the effects of placing a global gate of thickness $t_g$ atop the set of wires, separated by an insulating oxide layer of thickness $t_\text{ox}$.
Here, the wires, oxide layer, and global gate are taken to be very long along $\hat x$ and $\hat y$, to suppress edge effects in the region of interest.
The strains are computed, as described above, for a range system parameters, and the resulting conduction-band energies $\Delta E_c$ are plotted as function of lateral position in Figs.~\ref{FIG_MultiWire}c and d.
\red{The magnitude of the oscillations in these figures is consistent with recent experimental observations in gated devices \cite{CorleyWiciak2023}.
It is also interesting to note that similarly sized  fluctuations (perhaps a little smaller), are also observed near crosshatch features, whose characteristic feature size is on the order of microns, rather than hundreds of nanometers \cite{CorleyWiciak2023}.}

We first consider the case without a global gate, corresponding to $t_g = 0$, and a thin oxide layer of thickness $t_\text{ox} = 2~\T{nm}$, matching the oxide thickness of Fig.~\ref{FIG_SingleWire}. 
We refer to the resulting strain behavior as ``gate-driven,'' noting that very similar behavior is observed when the oxide is removed.
The computed $\Delta E_c$ results are shown as a gray line in Fig.~\ref{FIG_MultiWire}c.
In this multigate device, $\Delta E_c$ exhibits short-range variations that mirror the local gate structure, and a slowly varying envelope arising from long-range strain fields.
The trough-to-peak amplitudes are similar to those observed in Fig.~\ref{FIG_SingleWire}, indicating that the short-range features are mainly governed by the gate structure right above the 2DEG.
It is important to note that the amplitudes of these oscillations are large enough to produce unintentional dots.
For example, the series of dips along the gray line in Fig.~\ref{FIG_MultiWire}c are 3~meV deep, compared to orbital excitation energies of 1-3~meV in typical dots.
Such potential fluctuations are also large enough to affect exchange interactions, which are used to construct two-qubit gates. 
For example, it has been shown that a useful exchange interaction of strength $J\approx 10~\text{MHz}$ requires having a tunnel barrier between two Si/SiGe dots with a height less than 1~meV~\cite{Burkard2023}. 
If necessary, such strain-induced variations could potentially be compensated electrostatically; however, for complicated gate geometries, the competition between short- and long-range strain features could make this challenging. 
A desirable approach is therefore to compensate for some of the fluctuation through gate design, as discussed below.
Finally, we note that the envelope of the potential variations can also vary by several meV, even across a series of identical wires.
For example, in this geometry, there is a roughly 1~meV difference in $\Delta E_c$ between neighboring wires, which can be viewed as an effective detuning shift between quantum dots.

Next, we study the same five-gate geometry, now including  an additional global gate.
To provide a systematic understanding of the competing effects between the metal gates and the oxide between the gates, we first consider the limiting case of no oxide layer ($t_\text{ox}=0$) and a very thick top gate ($t_g \rightarrow\infty$). 
In this case, the two metal layers are contiguous and the nonuniform structural features on the top of the device (see Figs.~\ref{FIG_MultiWire}a, b) are so far away that they have no significant effect.
The resulting gate structure (not shown) is then effectively uniform, resulting in uniform strain in the $x$-$y$ plane.
Now if we consider the limit $t_g \rightarrow \infty$ with $t_\text{ox}>0$, any emerging strain fluctuations must be caused by the thin layer of oxide around the lower gates that replaces the metal in the previous geometry. 
An example of such fluctuations is given by the dashed black line in Fig.~\ref{FIG_MultiWire}c, for the case of $t_\text{ox} = 2~\T{nm}$ and $t_g = 700~\T{nm}$, where the latter approximates the limit $t_{g} \rightarrow \infty$.
We refer to the resulting fluctuations as ``oxide-driven.''
Remarkably, such oxide-driven behavior persists even down to typical $t_g$ values in modern devices. 
For example, the solid black line in Fig.~\ref{FIG_MultiWire}c shows results for a global gate with $t_g = 70~\T{nm}$, whose height matches the five Al wires, but whose behavior is similar to the $t_g \rightarrow \infty$ limit (dashed black line).
The oxide-driven fluctuations observed here are smaller in magnitude than the gate-driven fluctuations, and importantly, they have the opposite sign.
As discussed below, this sign change suggests that strain engineering principles, combined with simulations, may be used to realistically suppress the short-range strain fluctuations.

A comparison of the gray and black lines in Fig.~\ref{FIG_MultiWire}c shows that a global top gate can reduce the strain fluctuations caused by wires.
However, this trend is not universal, as observed in Fig.~\ref{FIG_MultiWire}d, where we consider the same gate geometries as Fig.~\ref{FIG_MultiWire}c, but with an increased oxide thickness of  $t_\text{ox} = 7~\T{nm}$.
The results show a similar sign change between the gate-driven and oxide-driven regimes; however, strain fluctuations under a thick gate (dashed line) are now stronger than the case with no global gate (gray line), and stronger than the fluctuations associated with the $t_g=70$~nm gate in Fig.~\ref{FIG_MultiWire}c,
\red{where the latter is a typical gate thickness, as considered in later sections.}
This behavior can be explained by the fact that the 70 nm gates in Figs.~\ref{FIG_MultiWire}c and d both fall into the oxide-driven regime, but the thicker oxide in Fig.~\ref{FIG_MultiWire}d drives this device deeper into the oxide-driven regime, where fluctuations are stronger.

Based on this discussion, we build the following intuition. A thick and uniform global gate, with no oxide layer between it and any buried patterned gates, results in a uniform strain distribution. This situation contrasts with the case of isolated patterned gates, which cause strain fluctuations. A thin oxide layer separating such isolated gates from a global gate, can therefore be viewed as ``negative space'' with respect to the metal, causing fluctuations of opposite sign as the isolated gates. Indeed, the thicker the oxide, the stronger the effect. Such gate-driven and oxide-driven behaviors compete, so an accurate description requires more-detailed simulations. However, the observed sign change is universal for the types of geometries studied here, indicating that careful design of global gate(s) and oxides may be able to strongly suppress the short-range fluctuations. 
In the current work, we demonstrate this by tuning $t_g$, for two different $t_\text{ox}$ values, as shown by the blue and pink lines in Figs.~\ref{FIG_MultiWire}c and d, respectively.
In both cases, we find that the short-range fluctuations are largely suppressed.
Moreover, in supplementary Fig.~S8, we show this suppression is robust to variations of $t_g$, over a range of several nanometers.
While long-range fluctuations in $\Delta E_c$ still persist across the gate array, we expect that such coarse features can be effectively compensated via electrostatic gating.
These results demonstrate the importance of strain simulations and device optimization for suppressing unwanted potential fluctuations.

\vspace{.2in}
\noindent \textbf{Varying the oxide thickness below realistic global gates} \\
In this section and the next, we perform strain simulations of realistic devices.
\red{Because of their complexity and large parameter spaces, we do not perform full optimizations of these device geometries, for example, to suppress strain fluctuations.}
However, in the current section, we vary the oxide thickness between a lower set of gates and a global overlapping gate, to explore the crossover between gate-driven and oxide-driven behaviors.
The Si/SiGe quantum-dot device studied in Ref.~\cite{Xue2021} provides a useful setting for this study, because the locally patterned (lower) gates are all formed in a single layer.

We consider two variations of the device used in Ref.~\cite{Xue2021}, as shown in Figs.~\ref{FIG_SingleLayer}a and b.
Here, the heterostructure consists of a 1.2~$\mu$m strain-relaxed 
Si$_{0.7}$Ge$_{0.3}$ buffer layer, an 8~nm Si quantum well, a 30~nm Si$_{0.7}$Ge$_{0.3}$ spacer, a 1~nm Si cap, a thin Al$_2$O$_3$ oxide layer, a set of 20~nm Al patterned metal gates, and another thin Al$_2$O$_3$  layer. 
(We also consider the effect of an additional cobalt micromagnet layer in supplementary Fig.~S6.)
To study the effects of oxide thickness, we consider two different cases: 7~nm oxide layers (consistent with the device in \cite{Xue2021}) and 2~nm oxide layers (consistent with the device in \cite{Neyens2019}), applied to the oxide layers above and below the lower gate layer.
This geometry is shown in Fig.~\ref{FIG_SingleLayer}a, while Fig.~\ref{FIG_SingleLayer}b also includes a 40~nm Al global gate.
Results for $\Delta E_c$ are shown in Figs.~\ref{FIG_SingleLayer}c-f, for the four different combinations of oxide thicknesses and gate geometries, with corresponding linecuts shown in Figs.~\ref{FIG_SingleLayer}g and h.

We study the effects of gate-layer stacking by comparing Figs.~\ref{FIG_SingleLayer}c and d.
For these two simulations, the oxide thickness is the same (2~nm), and the only difference is the absence or presence of a global top gate.
The linecut in Fig.~\ref{FIG_SingleLayer}g for the case with no global top gate (solid-purple line) shows strong short-range fluctuations, with a similar amplitude and shape as previous simulations, while the case with the stacked global gate (dashed-gray line) shows suppressed fluctuations.

The crossover between gate-driven and oxide-driven behavior can be seen most clearly by comparing Figs.~\ref{FIG_SingleLayer}c and f, and the corresponding linecuts in Fig.~\ref{FIG_SingleLayer}g.
Here, the solid purple linecut, with a thin oxide layer and no overlapping gate, clearly lies in the gate-driven regime.
In contrast, the dashed-black linecut describes a thick, globally overlapping gate with a relatively thick oxide, and exhibits fluctuation oscillations with similar magnitude and opposite sign, as consistent with the oxide-driven regime.
While we do not attempt to optimize the current device, we note that the dashed-gray linecut, with an overlapping gate but a thinner oxide layer, is intermediate between the gate-driven and oxide-driven regimes and exhibits strongly suppressed fluctuations.
Interestingly, the fourth linecut here (solid-gold line), corresponding to a thicker oxide with no overlapping gate, also exhibits suppressed fluctuations, although we do not explore this behavior further.

Finally, in Fig.~\ref{FIG_SingleLayer}h, we consider linecuts along the $\hat y$ axis that pass through the nominal centers of the quantum dots (green stars).
Here, we consider only the full gate geometry, which includes the global top gate.
Similar to linecuts along $\hat x$, these results confirm that stronger short-range fluctuations are obtained in the case of thicker oxides (gray curves).
We also note the slight differences between results in the left- and right-hand dots in Fig.~\ref{FIG_SingleLayer}h (solid vs dashed curves), which are caused by long-range strain fields arising from the asymmetric top-gate configuration.

\vspace{.4in}
\noindent \textbf{Strain fluctuations under overlapping gates} \\
We now study a second type of realistic gate design for quantum dot qubits, often simply referred to as ``overlapping gates.''
Here, we consider the quadruple quantum dot design of Ref.~~\cite{Neyens2019}, as illustrated in Figs.~\ref{FIG_Quad}a and b.
For the simulations, we assume the same heterostructure as in Fig.~\ref{FIG_SingleWire} --- now with three overlapping layers of Al metal gates.
From bottom to top, these correspond to screening gates (maroon), plunger or reservoir gates (blue), and tunnel-barrier gates (yellow). 
We also include 2~nm Al$_2$O$_3$ insulating layers between each of the metal layers.
(Note that thin oxide layers is a key requirement for effective overlapping gates, so we do not consider thicker oxide layers here.)
In the lower half of the images, the plunger gates define four quantum dots and two 2DEG reservoirs, on either side of the dots.
In the upper half of the images, the plungers define two charge-sensing dots, surrounded by three 2DEG reservoirs. 
To avoid edge effects, we use a full simulation cell of size  $18\times 18$~$\mu$m$^2$ in the $x$-$y$ plane and 2~$\mu$m in the $\hat z$ direction.
We have checked that this system size yields converged results.

The goal of these simulations is not to optimize the device but to characterize the strain features arising from overlapping gates. \red{To this end, we also do not explore the crossover between oxide- and gate-driven regimes.}
To disentangle the effects of the different gate layers, we begin with a smaller set of gates, and progressively add more gates until reaching a complete set.
Thus, Fig.~\ref{FIG_Quad}a includes only the screening, plunger, and upper-middle reservoir gates, while Fig.~\ref{FIG_Quad}b includes all the gates.
(Two intermediate gate sets are also considered in supplementary Fig.~S4.)
The results for the linecuts indicated in Figs.~\ref{FIG_Quad}c and d are shown in Figs.~\ref{FIG_Quad}e and f. 
The effectiveness of the gate-stacking strategy is seen most clearly by comparing the horizontal lines in Figs.~\ref{FIG_Quad}c and d.
In Fig.~\ref{FIG_Quad}c, this linecut crosses just the four plunger gates, which are separated by gaps.
In Fig.~\ref{FIG_Quad}d, the gaps are all filled by overlapping tunnel-barrier gates; reservoir gates are also present on either side, so the whole cut is covered by metal.
For the geometry with gaps between the gates (dot-dashed line in Fig.~\ref{FIG_Quad}e), we observe short-range oscillations with an amplitude of about 3~meV, similar to the isolated gates in Fig.~\ref{FIG_MultiWire}, while for the fully covered geometry (solid line), the short-range oscillations are suppressed by a factor of $\sim 3$.

Owing to the complexity of the device, these simulation results are more difficult to generalize.
However, the following observations may be useful.
First, since the plunger gates are not covered by tunnel-barrier gates in Figs.~\ref{FIG_Quad}a and c, the gray dot-dashed line in Fig.~\ref{FIG_Quad}e should exhibit gate-driven behavior, \red{as confirmed in our simulations.
Second, since the oxide thickness is only 2~nm, the solid-black line in Fig.~\ref{FIG_Quad}e could also potentially show gate-driven behavior, similar to the 2~nm oxides in previous sections. 
However, the resulting energy oscillations are found to be suppressed in our simulations --- neither gate-driven nor oxide-driven.}
Thus, although we do not perform device optimization \red{here, the full geometry already shows some suppression of the fluctuations.}
To suppress these fluctuations further, our previous simulations suggest several strategies: (i)
adding a global top gate, (ii) changing the thickness of the plunger and barrier gates, or (iii) changing the thickness of the oxide layers.
Additionally, we note that the tunnel-barrier gates in Fig.~\ref{FIG_Quad} are not globally overlapping -- they are only wide enough to fill the gaps between the plunger gates.
An alternative strategy could therefore be to modify the overlapping gate shapes, to provide more complete coverage. 

Finally, we note that strain from the screening gates can have unexpected effects on quantum-dot confinement in the $\hat y$ direction, as demonstrated in Fig.~\ref{FIG_Quad}f.
Here, the combination of screening and plunger gates produces \red{a dip feature on one side of the dot, which is weakly affected by the presence of the tunnel-barrier and reservoir gates.}
\red{The peak on the other side of the dot has the opposite sign as the electrostatic dot confinement.
To estimate its effect, we fit the peak to an upside-down confinement potential, $-m_t\omega_y^2y^2/2$, yielding a characteristic orbital energy of $\hbar\omega_y\approx 0.57$~meV,} where $m_t$ is the transverse effective mass of silicon.
Comparing this with typical dot orbital energies of $\hbar\omega=1$-3~meV (see supplementary Fig.~S5) suggests that the strain-induced effect is not likely to produce a double-dot along $\hat y$; however, it could affect the dot shape --- effectively flattening and widening the bottom of the confinement potential.
For weaker electrostatic confinements, strain-induced effects could potentially dominate, causing spurious dots to form.
Flattened dot potentials can also have a strong effect on singlet-triplet splitting due to electron-electron interactions~\cite{Ekmel2021}.
\vspace{0.15 in}

\vspace{.2in}
\noindent \textbf{\large Discussion} \\
We have investigated local strain fluctuations in realistic Si/SiGe quantum dot devices, arising from inhomogeneous materials layers --- in particular, from competing effects between metal gates and the oxide layer surrounding them.
To model the gates, we include strains from both thermal contractions and depositional stress, finding that the former dominates for Al gates, while the latter can dominate for Pd gates 
\red{and for Co micromagnets. (See supplementary Fig.~S6.)}
We note that local changes in the band-edge potential ($\Delta E_c$) arise mainly from the $\hat z$-axis strain ($\varepsilon_{zz}$), which is sensitive to 3D gate structure --- for example, from overlapping gates or oxide layers that fill the spaces between the gates.
Due to the complexity of the devices, the resulting strain patterns are not always intuitive. 

Despite this complexity, we can draw some general conclusions about the strain patterns.
First, we note that $\Delta E_c$ exhibits both short-range variations that mirror the shape of the top gates, and slowly varying envelopes that modulate the fast oscillations and may arise from edge effects.
The amplitude of the short-range fluctuations is comparable or smaller than typical electrostatic potentials arising from the same gates; however, unlike electrostatic potential variations, strain-induced effects cannot be modified after fabrication.
In many cases, the potential fluctuations can affect the dot shape, as seen in supplementary Fig.~S5.
Indeed, dot localization effects that were previously attributed to trapped charges~\cite{PhysRevB.80.075310,PhysRevB.80.115331,Hu2009} could potentially be explained, in full or in part, by strain effects.
Long-range fluctuations can also cause potential shifts below neighboring plunger gates that act like built-in detuning potentials between the dots, which must be electrostatically compensated for proper gate operation.

The most important result reported in this work is the observation that gate-layer stacking provides a key tool for modifying strain fluctuations.
Systematic simulations \red{in regular test structures} indicate two limiting types of behavior for overlapping gate geometries: gate-driven vs oxide-driven, where the former arises from the layer with isolated gates, while the latter arises when an oxide layer takes the place of metal in an otherwise contiguous gate, yielding strain of opposite sign.
Since this sign change can always be achieved by an appropriate choice of parameters, in an overlapping gate structure, it offers possibilities for suppressing strain fluctuations \red{in realistic devices} through careful design.
Indeed, the behavior observed in many devices can be explained within this general framework, including for devices with a global top gate (Fig.~\ref{FIG_SingleLayer}) and devices where narrow tunnel-barrier gates fill the gaps between plunger gates (Fig.~\ref{FIG_Quad}).

In future work, as arrays of qubits are scaled up, and every aspect of qubit functionality is tightly controlled, it will be increasingly important to account for strain-induced confinement effects, which can compete with electrostatic confinement.
The simplest approach to avoiding strain-induced fluctuations is to specifically design gate geometries to suppress them.
Gate-layer stacking and simulation-based optimization are key tools for achieving this goal. 
We expect strain engineering to play an important role in future qubit technologies.

\vspace{.2in}
\noindent \textbf{\large Methods} \\
\textbf{Strain simulations} \\
We calculate strain profiles using the Solid Mechanics module of COMSOL Multiphysics~\cite{Comsol}. To perform these calculations, we must specify the Young's modulus $E$, Poisson's ratio $\nu$, and coefficients of thermal expansion (CTE) $\alpha$, for each material. Note that each of these parameters varies with temperature, although COMSOL assumes temperature-independent values. $E$ and $\nu$ depend only weakly on the temperature, so we simply adopt room-temperature values for these parameters, from the COMSOL parameter library. To account for the temperature dependence of the CTEs, we define temperature-averaged CTEs as $\bar{\alpha} = (T_i - T_f)^{-1}\int_{T_f}^{T_i} \alpha(T)\, dT$, where $T_i$ and $T_f$ are the initial and final temperatures, respectively. We approximate these integrals using the temperature-dependent CTE data presented in Refs.~\cite{bradley2013, Simon1994, McSkimin1953,Levinshtein2001, Corruccini1961,Burke1962}. For the SiGe alloy, we use data for Si and Ge in combination with Vegard's law to evaluate its CTE. The resulting temperature-averaged CTEs used here are given by $\{\bar{\alpha}_{\text{Si}}, \bar{\alpha}_{\text{Si}_{0.7}\text{Ge}_{0.3}}, \bar{\alpha}_{\text{Al}}, \bar{\alpha}_{\text{Pd}}, \bar{\alpha}_{\text{Al}_{2}\text{O}_{3}}\} = \{0.76, 1.52, 14.16, 8.86, 3.30\}$ $(\times 10^{-6} \text{~K}^{-1})$. 

As described in supplementary Sec.~S2, biaxial strain in the quantum well is caused by lattice mismatch with the strain-relaxed SiGe barriers, and is given by $\varepsilon_{xx} = \varepsilon_{yy} = 1.2707\%$ and $\varepsilon_{zz} = -0.9798\%$. 
In the COMSOL simulations, we impose biaxial strain through initial-strain parameters.
Where specified, depositional stress in the metal gates is imposed through initial-stress parameters, which include the effects of thermal contraction as the metal is cooled from depositional temperatures to $293~\T{K}$.
We note that throughout this work, no depositional stress has been assumed for the Al$_2$O$_3$, because of the wide range of stress values reported in the literature (even sign changes), and their dependence on depositional tools and thermal postprocessing~\cite{Ylivaara2022}; however, we do not expect this approximation to affect our main results qualitatively, only quantitatively.
Finally, note that free boundary conditions~\cite{Comsol} are enforced on every surface, except for the bottom surface, where we enforce contraction that matches the bulk thermal contraction of Si, to emulate the presence of a semi-infinite Si substrate. 

The calculated strain profiles are related to the offsets $\Delta E_c$ of the conduction band in the Si quantum well through the relation~\cite{Thorbeck}
\begin{equation}
    \Delta E_c = E_0 + \Xi_u \varepsilon_{zz} + \Xi_d (\varepsilon_{xx} + \varepsilon_{yy} + \varepsilon_{zz}), \label{DeltaEc}
\end{equation}
where $\Xi_u = 10.5$~eV and $\Xi_d = 1.1$~eV are deformation potential coefficients~\cite{Fischetti1996}, whose difference is a consequence of the anisotropy of the $z$ valleys, which are the low-energy points of the conduction band in momentum space for the strained Si/SiGe quantum well~\cite{Zwanenburg2013}. As a result, strain effects in $\Delta E_c$ are dominated by the $z$ component of the strain tensor, $\varepsilon_{zz}$, as can be seen by comparing Figs.~\ref{FIG_SingleWire}c and d.
When computing energy fluctuations from Eq.~(\ref{DeltaEc}), we use the strain evaluated in the 2DEG plane, which is assumed to lie $1.5~\text{nm}$ below the top interface of the quantum well, where the low-energy electronic wave functions of the quantum well are concentrated. Finally, unless otherwise specified, $E_0$ is shifted in each calculation such that the asymptotic value of $\Delta E_c$ is zero, far away from the active region containing the gates.

Additional analytical methods, numerical methods, and model parameters \cite{Wortman, SiGe, Si&Ge2000, Si&GeRu, Auld1973, Stopa, Pletea2006, Graciano2019, Lieven, Barron1980} are discussed in the supplementary information.

\vspace{.2in}
\noindent \textbf{\large Data Availability} \\
The numerical data sets generated and analyzed in this work are available in the repository named ``Reducing strain fluctuations in quantum dot devices by gate-layer stacking'' at the following address:  (zenodo)

\vspace{.2in}
\noindent \textbf{\large Code Availability} \\
Simulations were performed using COMSOL Multiphysics version 6.4, making use of the CAD Import Module, the Solid Mechanics Interface in the Structural Mechanics Module, and the Electrostatics Interface in the AC/DC Module.
The scripts and COMSOL codes used to perform all numerical analyses in this work are available alongside the numerical data in the repository named ``Reducing strain fluctuations in quantum dot devices by gate-layer stacking'' at the following address:  (zenodo)

\vspace{.2in}
\noindent \textbf{\large Author Contributions} \\
C.D.F.\ and T.O.\ performed strain calculations with support from E.S.J.\ and E.R.M. 
M.P.L.\ performed electrostatic calculations. C.D.F., T.O., E.S.J., B.D.W., M.A.E., and M.F.\ analyzed the results and wrote the manuscript.

\vspace{.2in}
\noindent \textbf{\large Competing Interests} \\
The authors declare no competing financial or non-financial interests.

\vspace{.2in}
\noindent \textbf{\large Acknowledgments} \\
We are grateful to Xiao Xue, Nodar Samkharadze, Gertjan Eenink, Menno Veldhorst, and Lieven Vandersypen for sharing device designs related to this work.
We also thank Don Stone and Hudaiba Soomro for important clarifications related to the simulations, and Don Savage for many insightful discussions.

\vspace{.2in}
\noindent \textbf{\large Funding} \\
This research was sponsored in part by the Army Research Office under continuing Awards No.\ W911NF-17-1-0274 and W911NF-23-1-0110.
M.P.L.\ and B.D.W.\ were supported by fellowships from the Army Research Office under Award No.\ W911NF-22-1-0090.
T.O.\ was sponsored by the Netherlands Ministry of Defence under Award No.\ QuBits R23/009. 
The views, conclusions, and recommendations contained in this document are those of the authors and are not necessarily endorsed by nor should they be interpreted as representing the official policies, either expressed or implied, of the Army Research Office, the U.S. Government, or the Netherlands Ministry of Defence. The U.S. Government is authorized to reproduce and distribute reprints for U.S. Government purposes notwithstanding any copyright notation herein. The Netherlands Ministry of Defence is also authorized to reproduce and distribute reprints for Government purposes notwithstanding any copyright notation herein.

\vspace{.2in}
\noindent \textbf{\large Additional information} \\
\noindent \textbf{Correspondence}
should be addressed to Benjamin D. Woods.

\bibliography{References}

@article{Loss1998,
  title = {Quantum computation with quantum dots},
  author = {Loss, Daniel and DiVincenzo, David P.},
  journal = {Phys. Rev. A},
  volume = {57},
  issue = {1},
  pages = {120--126},
  numpages = {0},
  year = {1998},
  month = {Jan},
  publisher = {American Physical Society},
  doi = {10.1103/PhysRevA.57.120},
  url = {https://link.aps.org/doi/10.1103/PhysRevA.57.120}
}

@article{Ylivaara2022,
    author = {Ylivaara, Oili M. E. and Langner, Andreas and Ek, Satu and Malm, Jari and Julin, Jaakko and Laitinen, Mikko and Ali, Saima and Sintonen, Sakari and Lipsanen, Harri and Sajavaara, Timo and Puurunen, Riikka L.},
    title = {Thermomechanical properties of aluminum oxide thin films made by atomic layer deposition},
    journal = {Journal of Vacuum Science \& Technology A},
    volume = {40},
    number = {6},
    pages = {062414},
    year = {2022},
    month = {11},
    issn = {0734-2101},
    doi = {10.1116/6.0002095},
    url = {https://doi.org/10.1116/6.0002095}
}

@article{Uriel2023,
  title = {Hole-Spin Driving by Strain-Induced Spin-Orbit Interactions},
  author = {Abadillo-Uriel, Jos\'e Carlos and Rodr\'{\i}guez-Mena, Esteban A. and Martinez, Biel and Niquet, Yann-Michel},
  journal = {Phys. Rev. Lett.},
  volume = {131},
  issue = {9},
  pages = {097002},
  numpages = {7},
  year = {2023},
  month = {Sep},
  publisher = {American Physical Society},
  doi = {10.1103/PhysRevLett.131.097002},
  url = {https://link.aps.org/doi/10.1103/PhysRevLett.131.097002}
}

@article{Pateras2018,
author = {Pateras, Anastasios and Park, Joonkyu and Ahn, Youngjun and Tilka, Jack A. and Holt, Martin V. and Reichl, Christian and Wegscheider, Werner and Baart, Timothy A. and Dehollain, Juan Pablo and Mukhopadhyay, Uditendu and Vandersypen, Lieven M. K. and Evans, Paul G.},
title = {Mesoscopic Elastic Distortions in {GaAs} Quantum Dot Heterostructures},
journal = {Nano Letters},
volume = {18},
number = {5},
pages = {2780-2786},
year = {2018},
doi = {10.1021/acs.nanolett.7b04603},
URL = {https://doi.org/10.1021/acs.nanolett.7b04603}}

@ARTICLE{Wortman,
  author={Wortman, J. J. and Evans, R. A.},
  journal={J. Appl. Phys.}, 
  title={Young's Modulus, Shear Modulus, and {Poisson's} Ratio in Silicon and Germanium}, 
  year={1965},
  volume={36},
  number={1},
  pages={153-156},
  doi={10.1063/1.1713863}}

@ARTICLE{Hopcroft,
  author={Hopcroft, Matthew A. and Nix, William D. and Kenny, Thomas W.},
  journal={Journal of Microelectromechanical Systems}, 
  title={What is the {Young's} Modulus of Silicon?}, 
  year={2010},
  volume={19},
  number={2},
  pages={229-238},
  doi={10.1109/JMEMS.2009.2039697}}

@misc{Lieven,
  author       = {Lieven M. K. Vandersypen},
  howpublished = {Personal communication},
  year         = {2026}
}

@book{auld1973,
  title={Acoustic Fields and Waves in Solids},
  author={Auld, B.A.},
  volume={1},
  isbn={9780471037002},
  lccn={72008926},
  series={A Wiley-Interscience publication},
  url={https://books.google.com/books?id=1JMpAQAAMAAJ},
  year={1973},
  publisher={Wiley}
}

@article{Zwanenburg2013,
  title = {Silicon quantum electronics},
  author = {Zwanenburg, Floris A. and Dzurak, Andrew S. and Morello, Andrea and Simmons, Michelle Y. and Hollenberg, Lloyd C. L. and Klimeck, Gerhard and Rogge, Sven and Coppersmith, Susan N. and Eriksson, Mark A.},
  journal = {Rev. Mod. Phys.},
  volume = {85},
  issue = {3},
  pages = {961--1019},
  numpages = {0},
  year = {2013},
  month = {Jul},
  publisher = {American Physical Society},
  doi = {10.1103/RevModPhys.85.961},
  url = {https://link.aps.org/doi/10.1103/RevModPhys.85.961}
}

@article{Burkard2023,
  title = {Semiconductor spin qubits},
  author = {Burkard, Guido and Ladd, Thaddeus D. and Pan, Andrew and Nichol, John M. and Petta, Jason R.},
  journal = {Rev. Mod. Phys.},
  volume = {95},
  issue = {2},
  pages = {025003},
  numpages = {58},
  year = {2023},
  month = {Jun},
  publisher = {American Physical Society},
  doi = {10.1103/RevModPhys.95.025003},
  url = {https://link.aps.org/doi/10.1103/RevModPhys.95.025003}
}

@article{Xue2022,
   author = {Xiao Xue and Maximilian Russ and Nodar Samkharadze and Brennan Undseth and Amir Sammak and Giordano Scappucci and Lieven M. K. Vandersypen},
   doi = {10.1038/s41586-021-04273-w},
   issn = {0028-0836},
   issue = {7893},
   journal = {Nature},
   month = {1},
   pages = {343-347},
   title = {Quantum logic with spin qubits crossing the surface code threshold},
   volume = {601},
   year = {2022},
}

@article{Noiri2022,
   author = {Akito Noiri and Kenta Takeda and Takashi Nakajima and Takashi Kobayashi and Amir Sammak and Giordano Scappucci and Seigo Tarucha},
   doi = {10.1038/s41586-021-04182-y},
   issn = {0028-0836},
   issue = {7893},
   journal = {Nature},
   month = {1},
   pages = {338-342},
   title = {Fast universal quantum gate above the fault-tolerance threshold in silicon},
   volume = {601},
   year = {2022},
}

@article{Mills2021,
author = {Adam R. Mills  and Charles R. Guinn  and Michael J. Gullans  and Anthony J. Sigillito  and Mayer M. Feldman  and Erik Nielsen  and Jason R. Petta },
title = {Two-qubit silicon quantum processor with operation fidelity exceeding 99\%},
journal = {Science Advances},
volume = {8},
number = {14},
pages = {eabn5130},
year = {2022},
doi = {10.1126/sciadv.abn5130},
URL = {https://www.science.org/doi/abs/10.1126/sciadv.abn5130},
}

@article{Schaffler1997,
   author = {Friedrich Sch{\"a}ffler},
   doi = {10.1088/0268-1242/12/12/001},
   issn = {0268-1242},
   issue = {12},
   journal = {Semiconductor Science and Technology},
   month = {12},
   pages = {1515-1549},
   title = {High-mobility {Si} and {Ge} structures},
   volume = {12},
   year = {1997},
}

@article{Thorbeck,
   author = {Ted Thorbeck and Neil M. Zimmerman},
   doi = {10.1063/1.4928320},
   issn = {2158-3226},
   issue = {8},
   journal = {AIP Advances},
   month = {8},
   pages = {087107},
   title = {Formation of strain-induced quantum dots in gated semiconductor nanostructures},
   volume = {5},
   year = {2015},
}

@article{Stein2021,
   author = {Ryan M. Stein and Z. S. Barcikowski and S. J. Pookpanratana and J. M. Pomeroy and M. D. Stewart},
   doi = {10.1063/5.0061369},
   issn = {0021-8979},
   issue = {11},
   journal = {Journal of Applied Physics},
   month = {9},
   pages = {115102},
   title = {Alternatives to aluminum gates for silicon quantum devices: Defects and strain},
   volume = {130},
   year = {2021},
}

@article{Mooy2020,
   author = {Brian Chi Ho Mooy and Kuan Yen Tan and Nai Shyan Lai},
   doi = {10.3390/universe6040051},
   issn = {2218-1997},
   issue = {4},
   journal = {Universe},
   month = {4},
   pages = {51},
   title = {Comparison of Strain Effect between Aluminum and Palladium Gated {MOS} Quantum Dot Systems},
   volume = {6},
   year = {2020},
}

@article{Barron1980,
   author = {T.H.K. Barron and J.G. Collins and G.K. White},
   doi = {10.1080/00018738000101426},
   issn = {0001-8732},
   issue = {4},
   journal = {Advances in Physics},
   month = {8},
   pages = {609-730},
   title = {Thermal expansion of solids at low temperatures},
   volume = {29},
   year = {1980},
}

@article{Abadias2018,
   author = {Grégory Abadias and Eric Chason and Jozef Keckes and Marco Sebastiani and Gregory B. Thompson and Etienne Barthel and Gary L. Doll and Conal E. Murray and Chris H. Stoessel and Ludvik Martinu},
   doi = {10.1116/1.5011790},
   issn = {0734-2101},
   issue = {2},
   journal = {Journal of Vacuum Science \& Technology A: Vacuum, Surfaces, and Films},
   month = {3},
   pages = {020801},
   title = {Review Article: Stress in thin films and coatings: Current status, challenges, and prospects},
   volume = {36},
   year = {2018},
}

@book{bradley2013,
  author = {Peter Bradley and Ray Radebaugh},
  title = {Properties of Selected Materials at Cryogenic Temperatures},
  year = {2013},
  publisher = {CRC Press, Boca Raton, FL},
  url = {https://tsapps.nist.gov/publication/get_pdf.cfm?pub_id=913059}
}

@book{Levinshtein2001,
year = {2001},
author = {Levinshtein, M. E. and Rumyantsev, Sergey L. and Shur, Michael},
address = {New York},
booktitle = {Properties of advanced semiconductor materials: GaN, AlN, InN, BN, SiC, SiGe},
isbn = {0471358274},
publisher = {Wiley},
title = {Properties of advanced semiconductor materials : GaN, AlN, InN, BN, SiC, SiGe }
}

@book{Si&Ge2000, 
year = {2000},
author = {Levinshtein, M. E. and Rumyantsev Sergey L. and Shur, Michael}, 
address = {Singapore}, 
booktitle = {Handbook Series on Semiconductor Parameters Volume 1: Si, Ge, C (Diamond), GaAs, GaP, GaSb, InAs, InP, InSb}, 
isbn = {978-981-02-2934-4}, 
publisher = {World Scientific}, 
title = {Handbook Series on Semiconductor Parameters Volume 1: Si, Ge, C (Diamond), GaAs, GaP, GaSb, InAs, InP, InSb}}

@inproceedings{Corruccini1961,
  title={Thermal Expansion of Technical Solids at Low Temperatures: A Compilation From the Literature},
  author={Robert Joseph Corruccini and John Gniewek},
  year={1961}
}

@misc{Si&GeRu,
note   = {\url{https://www.ioffe.ru/SVA/NSM/introduction.html}},
year = {2003},
title = {Electronic archive:
New Semiconductor Materials. Characteristics and Properties}
}

@article{Afshar2010,
   author = {F. Afshar and S. Nazarpour and A. Cirera},
   doi = {10.1063/1.3505725},
   issn = {0021-8979},
   issue = {9},
   journal = {Journal of Applied Physics},
   month = {11},
   pages = {093513},
   title = {Survey of the theory and experimental measurements of residual stress in {Pd} thin film},
   volume = {108},
   year = {2010},
}

@article{Xue2021,
   author = {Xiao Xue and Bishnu Patra and Jeroen P. G. van Dijk and Nodar Samkharadze and Sushil Subramanian and Andrea Corna and Brian Paquelet Wuetz and Charles Jeon and Farhana Sheikh and Esdras Juarez-Hernandez and Brando Perez Esparza and Huzaifa Rampurawala and Brent Carlton and Surej Ravikumar and Carlos Nieva and Sungwon Kim and Hyung-Jin Lee and Amir Sammak and Giordano Scappucci and Menno Veldhorst and Fabio Sebastiano and Masoud Babaie and Stefano Pellerano and Edoardo Charbon and Lieven M. K. Vandersypen},
   doi = {10.1038/s41586-021-03469-4},
   issn = {0028-0836},
   issue = {7858},
   journal = {Nature},
   month = {5},
   pages = {205-210},
   title = {{CMOS}-based cryogenic control of silicon quantum circuits},
   volume = {593},
   year = {2021},
}

@article{Neyens2019,
   author = {Neyens, Samuel F. and MacQuarrie, E. R. and Dodson, J. P. and J. Corrigan and Nathan Holman and Brandur Thorgrimsson and M. Palma and Thomas McJunkin and Edge, L. F. and Mark Friesen and Coppersmith, S. N. and Eriksson, M. A.},
   doi = {10.1103/PhysRevApplied.12.064049},
   issn = {2331-7019},
   issue = {6},
   journal = {Physical Review Applied},
   month = {12},
   pages = {064049},
   title = {Measurements of Capacitive Coupling Within a Quadruple-Quantum-Dot Array},
   volume = {12},
   year = {2019},
}

@article{Stopa,
  title = {Quantum dot self-consistent electronic structure and the Coulomb blockade},
  author = {Stopa, M.},
  journal = {Phys. Rev. B},
  volume = {54},
  issue = {19},
  pages = {13767--13783},
  numpages = {0},
  year = {1996},
  month = {Nov},
  publisher = {American Physical Society},
  doi = {10.1103/PhysRevB.54.13767},
  url = {https://link.aps.org/doi/10.1103/PhysRevB.54.13767}
}

@inproceedings{Guisbiers2006,
   author = {G. Guisbiers and Olivier Van Overschelde and Steffen Strehle and M. Wautelet},
   doi = {10.1063/1.2173565},
   issn = {0094243X},
   journal = {AIP Conference Proceedings},
   pages = {317-324},
   publisher = {AIP},
   title = {Residual Stresses in {Ta, Mo, Al and Pd} Thin Films Deposited by E-Beam Evaporation Process on {Si and Si/SiO2} Substrates},
   year = {2006},
}

@article{Fischetti1996,
   author = {M. V. Fischetti and S. E. Laux},
   doi = {10.1063/1.363052},
   issn = {0021-8979},
   issue = {4},
   journal = {Journal of Applied Physics},
   month = {8},
   pages = {2234-2252},
   title = {Band structure, deformation potentials, and carrier mobility in strained {Si, Ge, and SiGe} alloys},
   volume = {80},
   year = {1996},
}

@misc{Comsol, 
title = {{COMSOL} Multiphysics\textsuperscript{\textregistered} v.~6.4. www.comsol.com. {COMSOL AB}, {Stockholm}, {Sweden}},
year={2025}
}

@article{McSkimin1953,
   author = {H. J. McSkimin},
   doi = {10.1121/1.1917687},
   issn = {0001-4966},
   issue = {4},
   journal = {The Journal of the Acoustical Society of America},
   month = {7},
   pages = {826-826},
   title = {Measurement of Elastic Constants at Low Temperatures by Means of High Frequency Ultrasonic Waves},
   volume = {25},
   year = {1953}
}

@article{Burke1962,
   author = {M. Burke},
   journal = {Journal of the American Ceramic Society},
   pages = {305-306},
   title = {Thermal expansion of ceramic materials at -200$^\circ$ to 0$^\circ$ {C}},
   volume = {45},
   year = {1962}
}

@article{Deng2018,
  title = {Negative exchange interactions in coupled few-electron quantum dots},
  author = {Deng, Kuangyin and Calderon-Vargas, F. A. and Mayhall, Nicholas J. and Barnes, Edwin},
  journal = {Phys. Rev. B},
  volume = {97},
  issue = {24},
  pages = {245301},
  numpages = {7},
  year = {2018},
  month = {Jun},
  publisher = {American Physical Society},
  doi = {10.1103/PhysRevB.97.245301},
  url = {https://link.aps.org/doi/10.1103/PhysRevB.97.245301}
}

@article{Deng2020,
  title = {Interplay of exchange and superexchange in triple quantum dots},
  author = {Deng, Kuangyin and Barnes, Edwin},
  journal = {Phys. Rev. B},
  volume = {102},
  issue = {3},
  pages = {035427},
  numpages = {14},
  year = {2020},
  month = {Jul},
  publisher = {American Physical Society},
  doi = {10.1103/PhysRevB.102.035427},
  url = {https://link.aps.org/doi/10.1103/PhysRevB.102.035427}
}

@article{Hu2006,
  title = {Charge-Fluctuation-Induced Dephasing of Exchange-Coupled Spin Qubits},
  author = {Hu, Xuedong and Das Sarma, S.},
  journal = {Phys. Rev. Lett.},
  volume = {96},
  issue = {10},
  pages = {100501},
  numpages = {4},
  year = {2006},
  month = {Mar},
  publisher = {American Physical Society},
  doi = {10.1103/PhysRevLett.96.100501},
  url = {https://link.aps.org/doi/10.1103/PhysRevLett.96.100501}
}

@article{SiGe,
author = {Goldberg, Y. and Levinshtein, Michael and Rumyantsev, Sergey},
year = {2001},
month = {01},
pages = {93-146},
title = {Properties of advanced semiconductor materials: {GaN, AIN, InN, BN, SiC, SiGe}},
volume = {25},
journal = {SciTech Book News}
}

@book{Simon1994,
title = {Cryogenic Properties of Inorganic Insulation Materials for ITER Magnets: A Review},
author = {Simon, N. J.},
publisher = {National Institute of Standards and Technology, Boulder, CO},
doi = {10.2172/761710},
url = {https://www.osti.gov/biblio/761710}, 
place = {United States},
year = {1994},
month = {12}
}

@article{Losert2023,
  title = {Practical strategies for enhancing the valley splitting in {Si/SiGe} quantum wells},
  author = {Losert, Merritt P. and Eriksson, M. A. and Joynt, Robert and Rahman, Rajib and Scappucci, Giordano and Coppersmith, Susan N. and Friesen, Mark},
  journal = {Phys. Rev. B},
  volume = {108},
  issue = {12},
  pages = {125405},
  numpages = {31},
  year = {2023},
  month = {Sep},
  publisher = {American Physical Society},
  doi = {10.1103/PhysRevB.108.125405},
  url = {https://link.aps.org/doi/10.1103/PhysRevB.108.125405}
}

@article{Hu2009,
  title = {Electron spin blockade and singlet-triplet transition in a silicon single electron transistor},
  author = {Hu, Binhui and Yang, C. H.},
  journal = {Phys. Rev. B},
  volume = {80},
  issue = {7},
  pages = {075310},
  numpages = {7},
  year = {2009},
  month = {Aug},
  publisher = {American Physical Society},
  doi = {10.1103/PhysRevB.80.075310},
  url = {https://link.aps.org/doi/10.1103/PhysRevB.80.075310}
}

@article{meyer2023b,
    author = {Meyer, Marcel and D\'eprez, Corentin and Meijer, Ilja N. and Unseld, Florian K. and Karwal, Saurabh and Sammak, Amir and Scappucci, Giordano and Vandersypen, Lieven M. K. and Veldhorst, Menno},
    title = {Single-Electron Occupation in Quantum Dot Arrays at Selectable Plunger Gate Voltage},
    journal = {Nano Letters},
    volume = {23},
    number = {24},
    pages = {11593-11600},
    year = {2023},
    month = {12},
    issn = {1530-6984},
    doi = {10.1021/acs.nanolett.3c03349},
    url = {https://doi.org/10.1021/acs.nanolett.3c03349}
}

@article{CorleyWiciak2023,
author = {Corley-Wiciak, Cedric and Richter, Carsten and Zoellner, Marvin H. and Zaitsev, Ignatii and Manganelli, Costanza L. and Zatterin, Edoardo and Schülli, Tobias U. and Corley-Wiciak, Agnieszka A. and Katzer, Jens and Reichmann, Felix and Klesse, Wolfgang M. and Hendrickx, Nico W. and Sammak, Amir and Veldhorst, Menno and Scappucci, Giordano and Virgilio, Michele and Capellini, Giovanni},
title = {Nanoscale Mapping of the {3D} Strain Tensor in a Germanium Quantum Well Hosting a Functional Spin Qubit Device},
journal = {ACS Applied Materials \& Interfaces},
volume = {15},
number = {2},
pages = {3119-3130},
year = {2023},
doi = {10.1021/acsami.2c17395},
URL = {https://doi.org/10.1021/acsami.2c17395}
}

@article{Wolfe:2024p034044,
  title = {Control of threshold voltages in $\mathrm{Si}$/$\mathrm{Si}$${}_{0.7}$$\mathrm{Ge}$${}_{0.3}$ quantum devices via optical illumination},
  author = {Wolfe, M.A. and Coe, Brighton X. and Edwards, Justin S. and Kovach, Tyler J. and McJunkin, Thomas and Harpt, Benjamin and Savage, D.E. and Lagally, M.G. and McDermott, R. and Friesen, Mark and Kolkowitz, Shimon and Eriksson, M.A.},
  journal = {Phys. Rev. Appl.},
  volume = {22},
  issue = {3},
  pages = {034044},
  numpages = {8},
  year = {2024},
  month = {Sep},
  publisher = {American Physical Society},
  doi = {10.1103/PhysRevApplied.22.034044},
  url = {https://link.aps.org/doi/10.1103/PhysRevApplied.22.034044}
}

@article{Zwerver2022,
  title = {Qubits made by advanced semiconductor manufacturing},
  author = {Zwerver, A. M. J. and Krähenmann, T. and Watson, T. F. and Lampert, L. and George, H. C. and Pillarisetty, R. and Bojarski, S. A. and Amin, P. and Amitonov, S. V. and Boter, J. M. and Caudillo, R. and Correas-Serrano, D. and Dehollain, J. P. and Droulers, G. and Henry, E. M. and Kotlyar, R. and Lodari, M. and Lüthi, F. and Michalak, D. J. and Mueller, B. K. and Neyens, S. and Roberts, J. and Samkharadze, N. and Zheng, G. and Zietz, O. K. and Scappucci, G. and Veldhorst, M. and Vandersypen, L. M. K. and Clarke, J. S.},
  journal = {Nature Electronics},
  volume = {5},
  issue = {3},
  pages = {184--190},
  numpages = {7},
  year = {2022},
  doi = {10.1038/s41928-022-00727-9},
  url = {https://doi.org/10.1038/s41928-022-00727-9}
}

@article{PaqueletWuetz2022,
   author = {Brian Paquelet Wuetz and Merritt P. Losert and Sebastian Koelling and Lucas E. A. Stehouwer and Anne-Marije J. Zwerver and Stephan G. J. Philips and Mateusz T. Mądzik and Xiao Xue and Guoji Zheng and Mario Lodari and Sergey V. Amitonov and Nodar Samkharadze and Amir Sammak and Lieven M. K. Vandersypen and Rajib Rahman and Susan N. Coppersmith and Oussama Moutanabbir and Mark Friesen and Giordano Scappucci},
   doi = {10.1038/s41467-022-35458-0},
   issn = {2041-1723},
   issue = {1},
   journal = {Nature Communications},
   month = {12},
   pages = {7730},
   title = {Atomic fluctuations lifting the energy degeneracy in {Si/SiGe} quantum dots},
   volume = {13},
   year = {2022},
}

@article{Zajac2016,
  title = {Scalable Gate Architecture for a One-Dimensional Array of Semiconductor Spin Qubits},
  author = {Zajac, D. M. and Hazard, T. M. and Mi, X. and Nielsen, E. and Petta, J. R.},
  journal = {Phys. Rev. Appl.},
  volume = {6},
  issue = {5},
  pages = {054013},
  numpages = {8},
  year = {2016},
  month = {Nov},
  publisher = {American Physical Society},
  doi = {10.1103/PhysRevApplied.6.054013},
  url = {https://link.aps.org/doi/10.1103/PhysRevApplied.6.054013}
}

@article{PRApplied044033,
  title = {Detuning Axis Pulsed Spectroscopy of Valley-Orbital States in $\mathrm{Si}$/$\mathrm{Si}$-$\mathrm{Ge}$ Quantum Dots},
  author = {Chen, Edward H. and Raach, Kate and Pan, Andrew and Kiselev, Andrey A. and Acuna, Edwin and Blumoff, Jacob Z. and Brecht, Teresa and Choi, Maxwell D. and Ha, Wonill and Hulbert, Daniel R. and Jura, Michael P. and Keating, Tyler E. and Noah, Ramsey and Sun, Bo and Thomas, Bryan J. and Borselli, Matthew G. and Jackson, C.A.C. and Rakher, Matthew T. and Ross, Richard S.},
  journal = {Phys. Rev. Appl.},
  volume = {15},
  issue = {4},
  pages = {044033},
  numpages = {13},
  year = {2021},
  month = {Apr},
  publisher = {American Physical Society},
  doi = {10.1103/PhysRevApplied.15.044033},
  url = {https://link.aps.org/doi/10.1103/PhysRevApplied.15.044033}
}

@article{Park2016,
    author = {Park, J. and Ahn, Y. and Tilka, J. A. and Sampson, K. C. and Savage, D. E. and Prance, J. R. and Simmons, C. B. and Lagally, M. G. and Coppersmith, S. N. and Eriksson, M. A. and Holt, M. V. and Evans, P. G.},
    title = "{Electrode-stress-induced nanoscale disorder in Si quantum electronic devices}",
    journal = {APL Materials},
    volume = {4},
    number = {6},
    pages = {066102},
    year = {2016},
    month = {06},
    doi = {10.1063/1.4954054},
    url = {https://doi.org/10.1063/1.4954054}
}

@article{Meyer2023,
author = {Meyer, Marcel and Déprez, Corentin and van Abswoude, Timo R. and Meijer, Ilja N. and Liu, Dingshan and Wang, Chien-An and Karwal, Saurabh and Oosterhout, Stefan and Borsoi, Francesco and Sammak, Amir and Hendrickx, Nico W. and Scappucci, Giordano and Veldhorst, Menno},
title = {Electrical Control of Uniformity in Quantum Dot Devices},
journal = {Nano Letters},
volume = {23},
number = {7},
pages = {2522-2529},
year = {2023},
doi = {10.1021/acs.nanolett.2c04446},
URL = {  
        https://doi.org/10.1021/acs.nanolett.2c04446
}
}

@ARTICLE{Massai2023,
  title    = "Impact of interface traps on charge noise and low-density transport properties in {Ge/SiGe} heterostructures",
  author   = "Massai, Leonardo and Het{\'e}nyi, Bence and Mergenthaler,
              Matthias and Schupp, Felix J and Sommer, Lisa and Paredes,
              Stephan and Bedell, Stephen W and Harvey-Collard, Patrick and
              Salis, Gian and Fuhrer, Andreas and Hendrickx, Nico W",
  journal  = "Communications Materials",
  volume   =  5,
  number   =  1,
  pages    = "151",
  month    =  aug,
  year     =  2024
}

@article{PhysRevB.80.075310,
  title = {Electron spin blockade and singlet-triplet transition in a silicon single electron transistor},
  author = {Hu, Binhui and Yang, C. H.},
  journal = {Phys. Rev. B},
  volume = {80},
  issue = {7},
  pages = {075310},
  numpages = {7},
  year = {2009},
  month = {Aug},
  publisher = {American Physical Society},
  doi = {10.1103/PhysRevB.80.075310},
  url = {https://link.aps.org/doi/10.1103/PhysRevB.80.075310}
}

@article{PhysRevB.80.115331,
  title = {Enhancement-mode double-top-gated metal-oxide-semiconductor nanostructures with tunable lateral geometry},
  author = {Nordberg, E. P. and Eyck, G. A. Ten and Stalford, H. L. and Muller, R. P. and Young, R. W. and Eng, K. and Tracy, L. A. and Childs, K. D. and Wendt, J. R. and Grubbs, R. K. and Stevens, J. and Lilly, M. P. and Eriksson, M. A. and Carroll, M. S.},
  journal = {Phys. Rev. B},
  volume = {80},
  issue = {11},
  pages = {115331},
  numpages = {10},
  year = {2009},
  month = {Sep},
  publisher = {American Physical Society},
  doi = {10.1103/PhysRevB.80.115331},
  url = {https://link.aps.org/doi/10.1103/PhysRevB.80.115331}
}

@article{Ekmel2021,
  title = {Strong electron-electron interactions in {Si/SiGe} quantum dots},
  author = {Ercan, H. Ekmel and Coppersmith, S. N. and Friesen, Mark},
  journal = {Phys. Rev. B},
  volume = {104},
  issue = {23},
  pages = {235302},
  numpages = {22},
  year = {2021},
  month = {Dec},
  publisher = {American Physical Society},
  doi = {10.1103/PhysRevB.104.235302},
  url = {https://link.aps.org/doi/10.1103/PhysRevB.104.235302}
}

@article{Liles2021,
  title = {Electrical control of the $g$ tensor of the first hole in a silicon MOS quantum dot},
  author = {Liles, S. D. and Martins, F. and Miserev, D. S. and Kiselev, A. A. and Thorvaldson, I. D. and Rendell, M. J. and Jin, I. K. and Hudson, F. E. and Veldhorst, M. and Itoh, K. M. and Sushkov, O. P. and Ladd, T. D. and Dzurak, A. S. and Hamilton, A. R.},
  journal = {Phys. Rev. B},
  volume = {104},
  issue = {23},
  pages = {235303},
  numpages = {9},
  year = {2021},
  month = {Dec},
  publisher = {American Physical Society},
  doi = {10.1103/PhysRevB.104.235303},
  url = {https://link.aps.org/doi/10.1103/PhysRevB.104.235303}
}

@article{Pletea2006, 
title = {In situ stress evolution of {Co} films sputtered onto oxidized {Si} (100) substrates},
author = {Pletea, M. and Brückner, W. and Wendrock, H. and Kaltofen, R. and Koch, R.},
journal = {Journal of Applied Physics},
  volume = {99},
  issue = {3},
  year = {2006},
  month = {Feb},
  publisher = {AIP Publishing},
  doi = {10.1063/1.2168243},
  url = {https://pubs.aip.org/aip/jap/article-abstract/99/3/033509/292712/In-situ-stress-evolution-of-Co-films-sputtered?redirectedFrom=fulltext}
}

@article{Graciano2019, 
title = {In-Situ Stress Measurements during Cobalt Electrodeposition},
author = {Graciano, V. P. and Bertocci, U. and Stafford, G. R.},
journal = {J. Electrochem. Soc.},
  volume = {166},
  issue = {1},
  year = {2019},
  month = {Jan},
  publisher = {ECS},
  doi = {10.1149/2.0311901jes},
  url = {https://iopscience.iop.org/article/10.1149/2.0311901jes}
}

\pagebreak
\clearpage

\renewcommand\thesection{S\arabic{section}}
\renewcommand\thefigure{S\arabic{figure}}
\renewcommand\theequation{S\arabic{equation}}
\renewcommand\thetable{S\arabic{table}}
\setcounter{figure}{0}
\setcounter{equation}{0}
\setcounter{section}{0}
\setcounter{table}{0}

\onecolumngrid
\vspace{0.5in}
\section*{Supplementary information}

In these supplementary sections, we provide details of the methods and results described in the main text.

\section{Anisotropic strain models for Si and SiGe}
\label{Sec:SMorthotropic}

For isotropic materials, a single Young's modulus $E$ relates the axial strain $\varepsilon$ to the stress $\sigma$ as $\varepsilon=\sigma/E$, while the negative ratio of the transverse strain to the axial strain is described by a single Poisson's ratio $\nu$.
For anisotropic materials, the stress-strain relation is generally replaced by a rank-4 tensor, taking the form
\begin{equation}
\varepsilon_{ij} = s_{ijkl} \sigma_{kl} .
\end{equation}
For cubic materials, however, the compliance tensor can be simplified, taking the form of a $6\times 6$ matrix:
\begin{equation}
\begin{pmatrix}
\varepsilon_{11} \\
\varepsilon_{22} \\
\varepsilon_{33} \\
\varepsilon_{23} \\
\varepsilon_{31} \\
\varepsilon_{12}
\end{pmatrix}
=
\begin{pmatrix}
s_{11} & s_{12} & s_{13} & 0      & 0      & 0 \\
s_{21} & s_{22} & s_{23} & 0      & 0      & 0 \\
s_{31} & s_{32} & s_{33} & 0      & 0      & 0 \\
0      & 0      & 0      & s_{44} & 0      & 0 \\
0      & 0      & 0      & 0      & s_{55} & 0 \\
0      & 0      & 0      & 0      & 0      & s_{66}
\end{pmatrix}
\begin{pmatrix}
\sigma_{11} \\
\sigma_{22} \\
\sigma_{33} \\
\sigma_{23} \\
\sigma_{31} \\
\sigma_{12}
\end{pmatrix} ,
\end{equation}
where the indices 1, 2, and 3 are often used synonymously with $x$, $y$, and $z$.
(Indeed, we use the $x$, $y$, and $z$ indices in the main text.)
The compliance coefficients are related to elastic constants as follows:
\begin{equation}
S=
\begin{pmatrix}
s_{11} & s_{12} & s_{13} & 0      & 0      & 0 \\
s_{21} & s_{22} & s_{23} &
0      & 0      & 0 \\
s_{31} & s_{32} & s_{33} & 0      & 0      & 0 \\
0      & 0      & 0      & s_{44} & 0      & 0 \\
0      & 0      & 0      & 0      & s_{55} & 0 \\
0      & 0      & 0      & 0      & 0      & s_{66}
\end{pmatrix}
=
\begin{pmatrix}
\frac{1}{E_{11}} & \frac{-\nu_{21}}{E_{22}} & \frac{-\nu_{31}}{E_{33}} & 0      & 0      & 0 \\
\frac{-\nu_{12}}{E_{11}} & \frac{1}{E_{22}} & \frac{-\nu_{32}}{E_{33}} & 0      & 0      & 0 \\
\frac{-\nu_{13}}{E_{11}} & \frac{-\nu_{23}}{E_{22}} & \frac{1}{E_{33}} & 0      & 0      & 0 \\
0      & 0      & 0      & \frac{1}{G_{23}} & 0      & 0 \\
0      & 0      & 0      & 0      & \frac{1}{G_{31}} & 0 \\
0      & 0      & 0      & 0      & 0      & \frac{1}{G_{12}}
\end{pmatrix}, \label{eq:compliance}
\end{equation}
where $E_{ij}$ are anisotropic Young's modulii, $\nu_{ij}$ are anisotropic Poisson's ratios, and $G_{ij}$ are anisotropic shear modulii.

It is well established in the literature that Si and Ge are anisotropic~\cite{Wortman}, and specifically, orthotropic. 
This form of anisotropy arises from the zinc-blende crystal structure. 
For a Si wafer oriented along [001], the in-plane directions [110] and $[\bar{1}10]$ are defined as `hard' directions, while the out-of-plane [001] direction, which corresponds to the direction of growth, is referred to as `soft.'
In this work, we define the $\hat x$, $\hat y$, and $\hat z$ directions as $[110]$, $[\bar{1}10]$, and $[001]$, respectively.
With these definitions, the room-temperature elastic constants for Si, Si$_{0.7}$Ge$_{0.3}$, and Ge are then given in Tables~S1-S3~\cite{Wortman, SiGe, Levinshtein2001, Si&Ge2000, Si&GeRu}.
Note that we perform a simple linear interpolation of the Si and Ge elastic constants to find the corresponding Si$_{0.7}$Ge$_{0.3}$ material parameters, before computing the compliance coefficients, as in Eq.~(\ref{eq:compliance}). 
Also note that these elastic properties show only a weak temperature dependence~\cite{Hopcroft};
we therefore adopt the reported room-temperature values for our simulations. 

\red{Based on these values, the stiffness matrices for Si, Ge, and Si$_{0.7}$Ge$_{0.3}$, oriented along [100], [010], and [001] are given by
\begin{equation}
C_\text{Si}
=
\begin{pmatrix}
165.7 & 63.9 & 63.9 & 0      & 0      & 0 \\
63.9 & 165.7 & 63.9 & 0      & 0      & 0 \\
63.9 & 63.9 & 165.7 & 0      & 0      & 0 \\
0      & 0      & 0      & 79.6 & 0      & 0 \\
0      & 0      & 0      & 0      & 79.6& 0 \\
0      & 0      & 0      & 0      & 0      & 79.6
\end{pmatrix}
\text{(GPa)} ,
\end{equation}
\begin{equation}
C_\text{Ge}
=
\begin{pmatrix}
129.2 & 47.9 & 47.9 & 0      & 0      & 0 \\
47.9 & 129.2 & 47.9 & 0      & 0      & 0 \\
47.9 & 47.9 & 129.2 & 0      & 0      & 0 \\
0      & 0      & 0      & 67 & 0      & 0 \\
0      & 0      & 0      & 0      & 67& 0 \\
0      & 0      & 0      & 0      & 0      & 67
\end{pmatrix}
\text{(GPa)} ,
\end{equation}
\begin{equation}
C_\text{SiGe}
=
\begin{pmatrix}
154.75 & 59.1 & 59.1 & 0      & 0      & 0 \\
59.1 & 154.75 & 59.1 & 0      & 0      & 0 \\
59.1 & 59.1& 154.75 & 0      & 0      & 0 \\
0      & 0      & 0      & 75.82 & 0      & 0 \\
0      & 0      & 0      & 0      & 75.82& 0 \\
0      & 0      & 0      & 0      & 0      & 75.82
\end{pmatrix}
\text{(GPa)} .
\end{equation}
These tensors are then rotated by $\pi/4$ to obtain the appropriate tensors in the $\{[110], [\bar{1}10], [001]\}$ frame, using the rotation $R \sigma R^T = \sigma'$, where $R$ is the rotation matrix given by \cite{Auld1973}}
\red{
\begin{equation}
\begin{pmatrix}
\cos^2(\theta) & \sin^2(\theta) & 0 & 0 & 0 & \sin(2\theta) \\
-\sin^2(\theta) & \cos^2(\theta) & 0 & 0 & 0 & -\sin(2\theta) \\
0 & 0 & 1 & 0 & 0 & 0 \\
0 & 0 & 0 & \cos(\theta) & -\sin(\theta) & 0 \\
0 & 0 & 0 & \sin(\theta) & \cos(\theta) & 0 \\
-\frac{\sin(2\theta)}{2} & \frac{\sin(2\theta)}{2} & 0 & 0 & 0 & \cos(2\theta)
\end{pmatrix} .
\end{equation}}
\red{In the rotated frame, the stiffness tensors for Si, Ge, and SiGe are then given by
\begin{equation}
C_\text{Si}
=
\begin{pmatrix}
194.4 & 35.2 & 63.9 & 0      & 0      & 0 \\
35.2 & 194.4 & 63.9 & 0      & 0      & 0 \\
63.9 & 63.9 & 165.7 & 0      & 0      & 0 \\
0      & 0      & 0      & 79.6 & 0      & 0 \\
0      & 0      & 0      & 0      & 79.6& 0 \\
0      & 0      & 0      & 0      & 0      & 50.9
\end{pmatrix}
\text{(GPa)} ,
\end{equation}
\begin{equation}
C_\text{Ge}
=
\begin{pmatrix}
155.55 & 21.55 & 47.9 & 0      & 0      & 0 \\
21.55 & 155.55 & 47.9 & 0      & 0      & 0 \\
47.9 & 47.9 & 129.2 & 0      & 0      & 0 \\
0      & 0      & 0      & 67 & 0      & 0 \\
0      & 0      & 0      & 0      & 67& 0 \\
0      & 0      & 0      & 0      & 0      & 40.65
\end{pmatrix}
\text{(GPa)} ,
\end{equation}
\begin{equation}
C_\text{SiGe}
=
\begin{pmatrix}
182.745 & 31.105 & 59.1 & 0      & 0      & 0 \\
31.105 & 182.745 & 59.1 & 0      & 0      & 0 \\
59.1 & 59.1& 154.75 & 0      & 0      & 0 \\
0      & 0      & 0      & 75.82 & 0      & 0 \\
0      & 0      & 0      & 0      & 75.82& 0 \\
0      & 0      & 0      & 0      & 0      & 47.825
\end{pmatrix}
\text{(GPa)} .
\end{equation}
These values are entered into our COMSOL Multiphysics simulations as an anisotropic linear elastic material, overriding the default compliance matrix, which is isotropic.}


\begin{figure}[b]
\begin{center}
\includegraphics[width=3.5in]{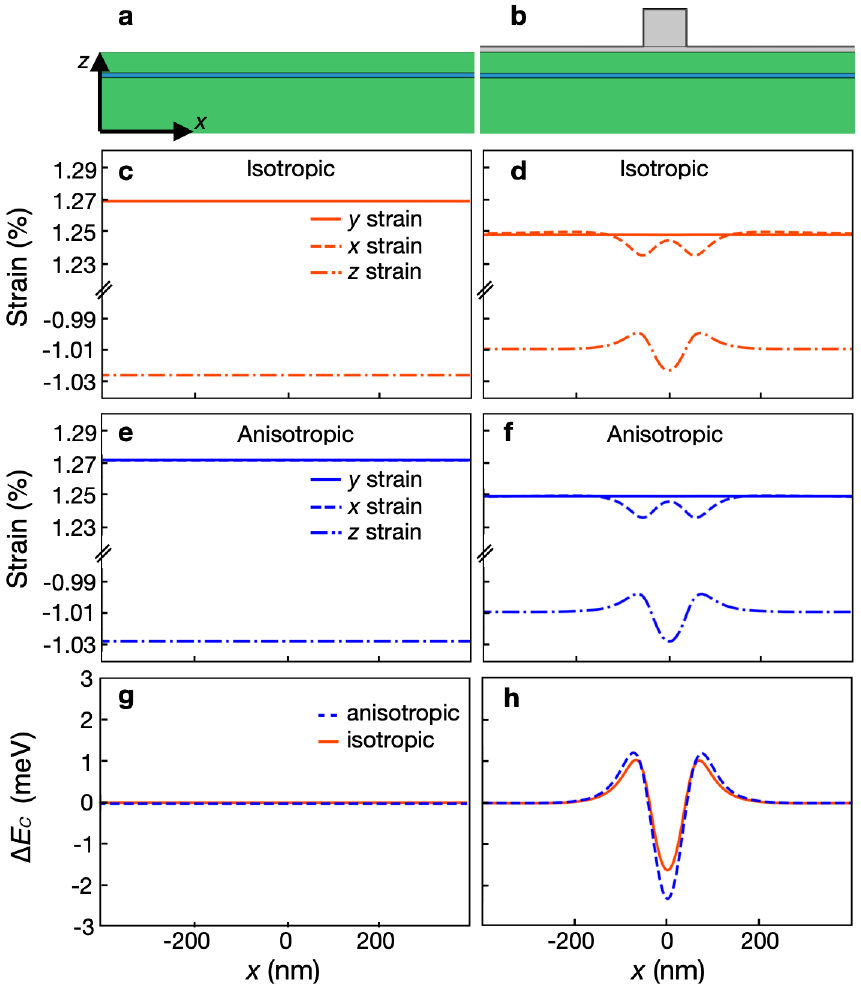}
\end{center}
\vspace{-1cm}
\hspace{-1cm}
\caption{
\textbf{Simulations comparing isotropic vs orthotropic strain models, as described in Sec.~S1.}
\textbf{a} Cross-section view of a materials stack, from bottom to top, comprising a Si$_{0.7}$Ge$_{0.3}$ virtual substrate (green), a Si quantum well (blue), and a Si$_{0.7}$Ge$_{0.3}$ spacer layer (green), with materials thicknesses of 2~$\mu$m, 9~nm, and 40~nm, respectively. 
\textbf{b} Same as \textbf{a} but with an additional insulating Al$_2$O$_3$ layer (light gray) of height 10~nm and a single Al wire (dark gray) of height \red{70~nm} and width 80~nm, oriented along $\hat y$. The metal wire is covered on the top and sides by a thin 2~nm Al$_2$O$_3$ layer (not shown for clarity). \textbf{c}-\textbf{f} Strain profiles for the geometry without the wire (left), and with the wire (right), obtained with the isotropic or orthotropic strain models, as indicated.  
\textbf{g}, \textbf{h} Potential energy fluctuations computed along a linecut through the quantum well, in the $\hat x$ direction.
The isotropic and orthotropic results are plotted in red and blue, respectively, as indicated.
Note that all results are obtained for thermally contracted systems at a temperature of 1~K.}
\label{FIG_Appendix_OrthovsIso}
\end{figure}

\begin{table}[t!]
\centering

\begin{minipage}[t]{0.32\textwidth}
\caption{Isotropic vs orthotropic properties of Si~\cite{SiGe, Levinshtein2001, Si&Ge2000, Si&GeRu}.
Note that the isotropic properties are consistent with the default values of COMSOL Multiphysics.}
\centering
\begin{tabular}{|c|c|c|} \hline
Property & Iso value & Ortho value \\ \hline
$E_{11} = E_{22}$ & 170 GPa & 169 GPa \\
$E_{33}$       & 170 GPa & 130 GPa \\
$G_{23}=G_{31}$ & 52 GPa & 80 GPa \\
$G_{12}$        & 52 GPa & 50 GPa \\
$\nu_{13}=\nu_{23}$ & 0.28 & 0.36 \\
$\nu_{31}=\nu_{32}$ & 0.28 & 0.28 \\
$\nu_{12}=\nu_{21}$ & 0.28 & 0.065 \\ \hline
\end{tabular}
\end{minipage}
\hfill
\begin{minipage}[t]{0.32\textwidth}
\caption{Isotropic vs orthotropic properties of Ge~\cite{SiGe, Levinshtein2001, Si&Ge2000, Si&GeRu}.
Note that the isotropic properties are consistent with the default values of COMSOL Multiphysics.}
\centering
\begin{tabular}{|c|c|c|} \hline
Property & Iso value & Ortho value \\ \hline
$E_{11} = E_{22}$ & 103 GPa & 138 GPa \\
$E_{33}$       & 103 GPa & 103 GPa \\
$G_{23}=G_{31}$ & 41 GPa & 65 GPa \\
$G_{12}$        & 41 GPa & 40 GPa \\
$\nu_{13}=\nu_{23}$ & 0.26 & 0.365 \\
$\nu_{31}=\nu_{32}$ & 0.26 & 0.27 \\
$\nu_{12}=\nu_{21}$ & 0.26 & 0.03 \\ \hline
\end{tabular}
\end{minipage}
\hfill
\begin{minipage}[t]{0.32\textwidth}
\caption{The isotropic and orthotropic properties of the Si$_{0.7}$Ge$_{0.3}$ Alloy are obtained via simple linear interpolation of the values presented in Tables~S1 and S2.}
\centering
\begin{tabular}{|c|c|c|} \hline
Property & Iso value & Ortho value \\ \hline
$E_{11} = E_{22}$ & 149.9 & 159.7 GPa \\
$E_{33}$       & 149.9 & 121.9 GPa \\
$G_{23}=G_{31}$ & 48.7 & 75.5 GPa \\
$G_{12}$        & 48.7 & 47 GPa \\
$\nu_{13}=\nu_{23}$ & 0.274 & 0.3615 \\
$\nu_{31}=\nu_{32}$ & 0.274 & 0.277 \\
$\nu_{12}=\nu_{21}$ & 0.274 & 0.0545 \\ \hline
\end{tabular}
\end{minipage}

\end{table}

To illustrate the quantitative differences arising from the isotropic vs orthotropic strain models, we have simulated the two simple wire geometries shown in Fig.~S1.
Here the left-hand column corresponds to a completely isotropic geometry with no wire, while in the right-hand column, the symmetry is broken, with a wire extending along $\hat y$.
The first set of results in panels c and d show results based on the isotropic strain model, while the next panels, e and f, show results from the orthotropic strain model.
The final two panels, g and h, show the computed energy variations.
Comparing panels c and e, we observe almost no difference in behaviors between isotropic and orthotropic, as expected for an isotropic geometry.
In panels d and f, the differences are also difficult to see, but they yield slightly different results for the energy fluctuations as shown in panel h.
To highlight these effects, in Fig.~S2, we plot the differences between isotropic and orthotropic results, for the strain and energy calculations.
We see that strain in the $\hat z$ direction is especially affected by the model parameters, yielding fluctuations that are \red{$>$24\%} larger for the orthotropic model, compared to the isotropic model.
This is important, because the energy fluctuations depend most strongly on the $z$ strain, giving total energy variations greater than 0.6~meV.

\begin{figure}[t]
\begin{center}
\includegraphics[width=3.in]{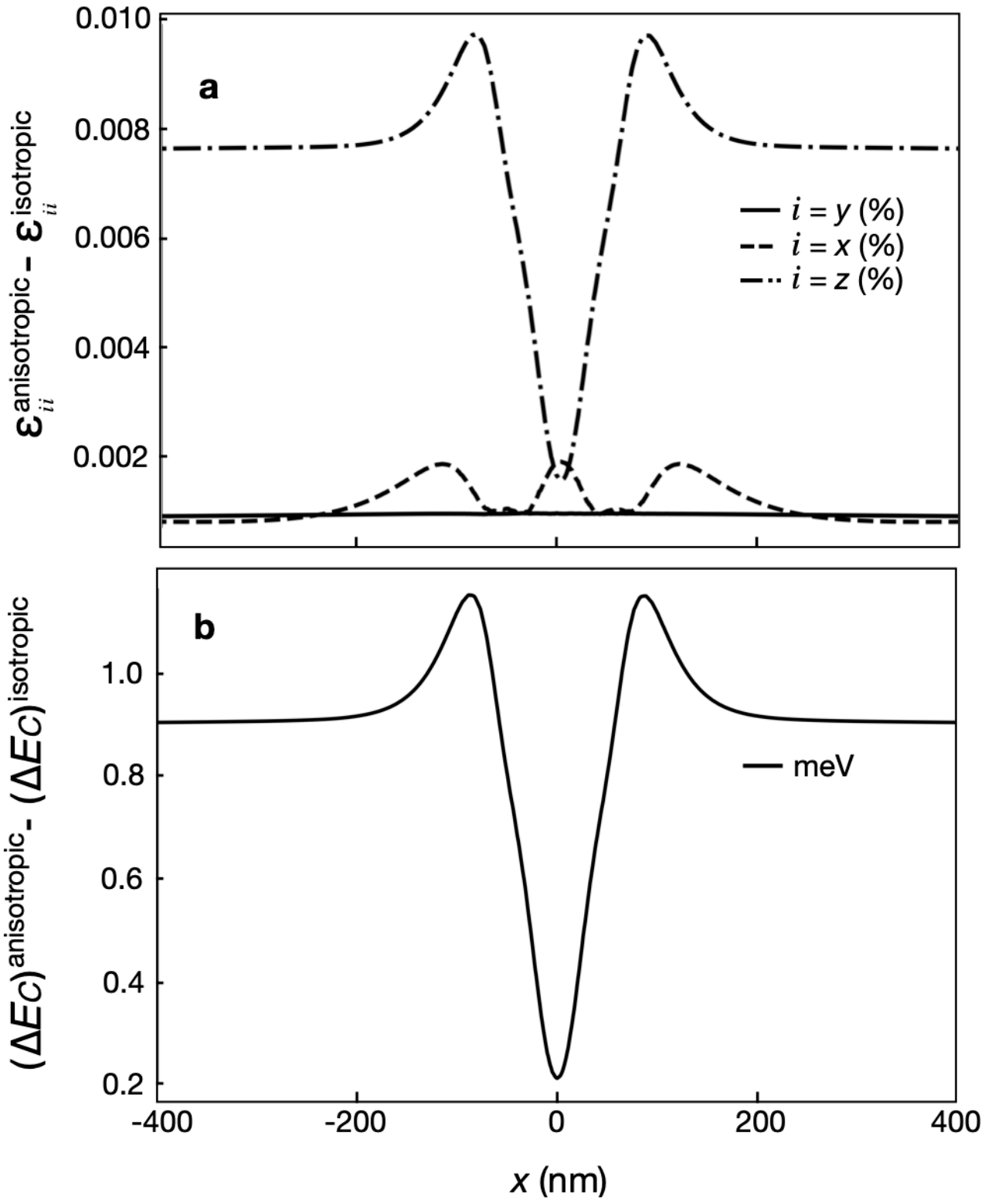}
\end{center}
\vspace{-1cm}
\hspace{-1cm}
\caption{
\textbf{Additional comparisons between isotropic and orthotropic strain models, for the device shown in Fig.~S1b.}
\textbf{a} Differences between strain results obtained from isotropic vs orthotropic strain models. 
We observe \red{$7.05 \%$} smaller fluctuations for the $x$ strain in the orthotropic model, compared to the isotropic model. Since the wire is oriented along the $\hat y$ axis, the $y$ fluctuations are within the numerical error of the simulation. 
The corresponding fluctuations of the $z$ strain are \red{$24.6 \%$} larger for the orthotropic model, compared to the isotropic model.
\textbf{b} Difference between potential energy fluctuations in isotropic vs orthotropic strain models.
We observe $21.5 \%$ larger fluctuations in the orthotropic model, compared to the isotropic model.  }
\label{FIG_Appendix_S02}
\end{figure}

\section{Biaxial strain in the quantum well due to lattice mismatch}
\label{sec:biaxial}

As noted in the Methods section of the main text, the biaxial strain in the quantum well, caused by lattice mismatch between the strain-relaxed SiGe virtual substrate and the strained Si quantum well, is included in our COMSOL simulations by imposing initial strain conditions.
In this supplementary section, we derive these initial conditions.

Since there is no initial shear strain, the shear components of the stress-strain relationship may be excluded in this discussion, yielding the simplified relation
\begin{equation}
\mathbf{}
\begin{pmatrix}
\varepsilon_{11} \\
\varepsilon_{22} \\
\varepsilon_{33} \\
\end{pmatrix}
=
\begin{pmatrix}
s_{11} & s_{12} & s_{13}\\
s_{21} & s_{22} & s_{23}\\
s_{31} & s_{32} & s_{33}\\
\end{pmatrix}
\begin{pmatrix}
\sigma_{11} \\
\sigma_{22} \\
\sigma_{33} \\
\end{pmatrix} .
\end{equation}
Due to the in-plane isotropy of the initial conditions, we may set $\sigma_{11}=\sigma_{22}=\sigma$ and $\sigma_{33}=0$.
Additionally, we define $\varepsilon_{11}=\varepsilon_{22}=\varepsilon_m$, where $\varepsilon_m$ is the lattice mismatch strain, given by
\begin{equation}
    \varepsilon_m=\frac{a_\text{SiGe}-a_\text{Si}}{a_\text{Si}}=0.01271 .
\end{equation}
Here, $a_\text{Si}=0.5431$~nm is the cubic lattice constant of Si, and $a_\text{SiGe}=0.5500$~nm is the lattice constant of the Si$_{0.7}$Ge$_{0.3}$ virtual crystal, obtained via Vegard's law. 
This accounts for the first two components of the initial strain tensor.
The final component, $\varepsilon_{33}$, can be determined as a response to the strain in the $\hat x$ and $\hat y$ directions.
Using the relations given above, and the compliance coefficients of Sec.~S1, we obtain
\begin{equation}
\varepsilon_{33}=\frac{2s_{31}}{s_{11}+s_{12}}\varepsilon_{m}=-0.00977 .
\end{equation}

\clearpage

\section{Supporting Figures}

In this section, we provide supplementary Figs.~S3-S10 to clarify discussions in the main text and provide supporting results.

\begin{figure}[h]
\begin{center}
\includegraphics[width=4in]{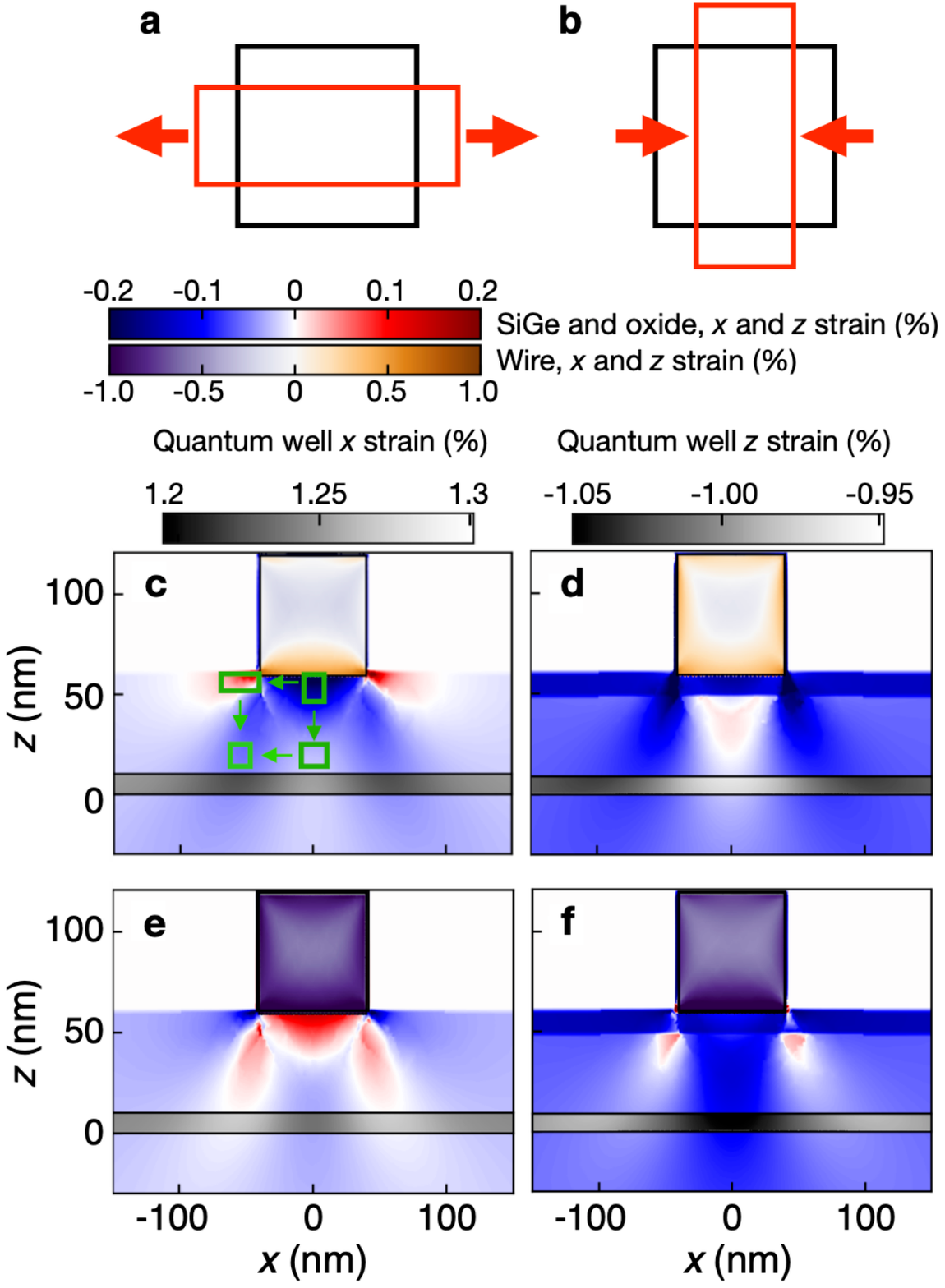}
\end{center}
\vspace{-0.5cm}
\hspace{-1cm}
\caption{
\textbf{Propagation of strain in a single-wire geometry.} 
\textbf{a} In-plane tensile stress deforms an unstrained square (black) into a rectangle (red) elongated in the horizontal direction and compressed in the vertical direction. 
\textbf{b} In-plane compressive stress deforms the square into a rectangle compressed in the horizontal direction and elongated in the vertical direction. 
\textbf{c} The $\varepsilon_{xx}$ strain profile for a Pd wire with $560~\text{MPa}$ of depositional tensile stress. 
Here, three different color scales are used to show the large strain variations across multiple materials. 
The green boxes illustrate the propagation of strain across the SiGe and oxide layers. 
Compressive strain is induced directly below the wire, causing deformations similar to \textbf{b}. 
In turn, this induces tensile strain on either side of the wire, causing deformations similar to \textbf{a}. 
Similar considerations lead to the deformations shown in \red{\textbf{d} and \textbf{f}}. \red{\textbf{e}} Same as \textbf{c}, except the Pd wire has $890~\text{MPa}$ of depositional compressive stress, causing a reversal of the strain pattern, as compared to \textbf{c}. \red{\textbf{d}}, \textbf{f} $\varepsilon_{zz}$ strain profiles, for the same conditions as \textbf{c} and \red{\textbf{e}}, respectively.}
\label{FIG_Appendix_Strains}
\vspace{-1mm}
\end{figure}

\begin{figure*}[t]
\begin{center}
\includegraphics[width=6.5in]{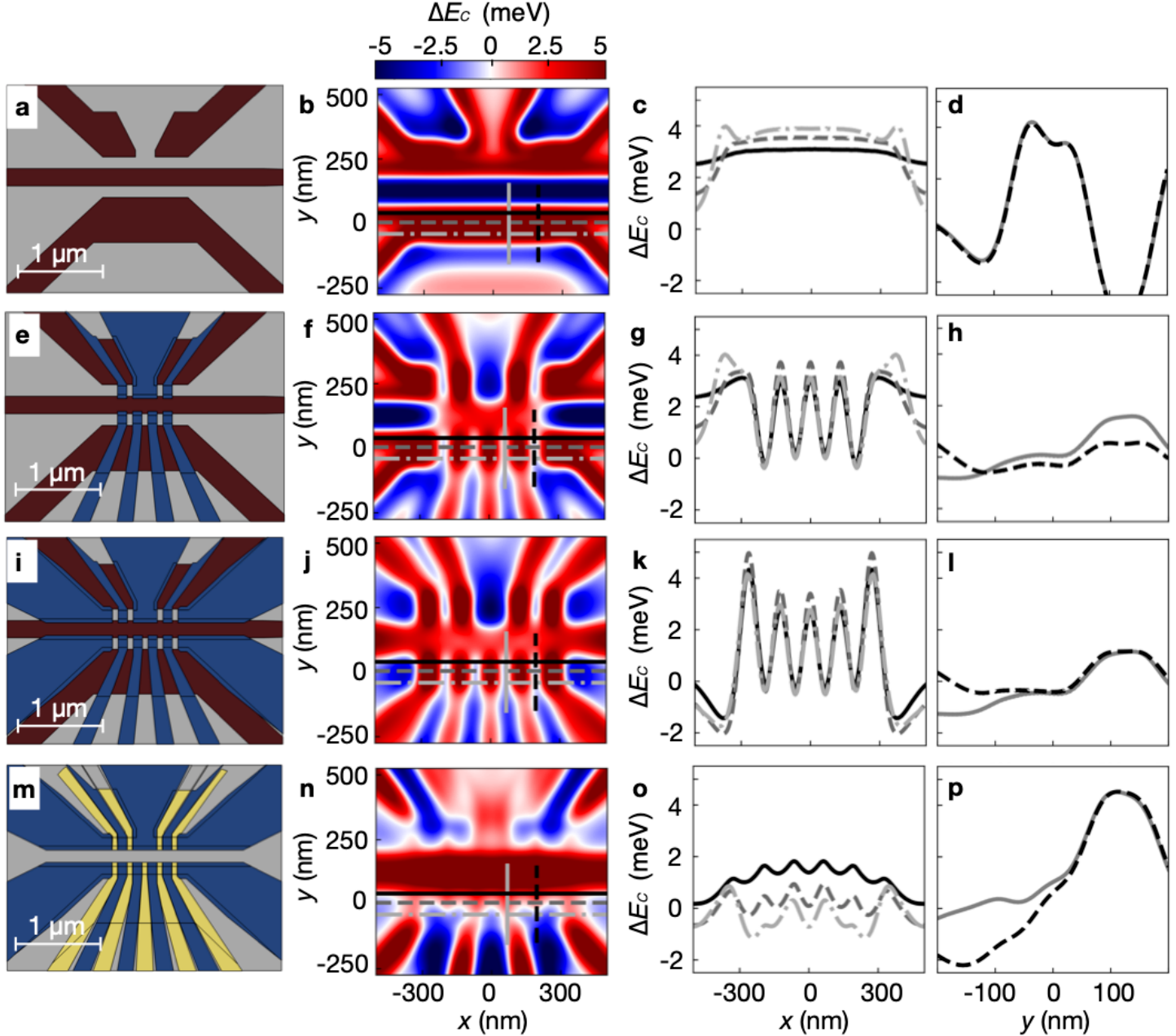}
\end{center}
\vspace{-0.5cm}
\hspace{-1cm}
\caption{
\textbf{Additional strain simulations of the quadruple-quantum-dot device shown in Fig.~5.}
The leftmost column shows the schematic gate layouts for each of the simulations. 
The second column shows the corresponding $\Delta E_c$ profiles, evaluated in the 2DEG plane. 
The third and fourth columns show horizontal and vertical linecuts, respectively, as indicated in the second column, with matching colors and line styles. 
Note that the horizontal linecuts pass through the lower four dots, while the vertical linecuts run through the centers of the third and fourth dots.
\textbf{a} The first gate set includes only the screening gates (maroon).
\textbf{b-d} A shallow potential well forms between the screening gates, as shown in \textbf{d}, while tunnel barriers form under the edges of the gates, similar to behavior observed in Fig.~3. 
\textbf{e} The same gate geometry studied in Fig.~5a, which includes plunger gates and the center-reservoir gate on the top portion of the device. 
\textbf{f-g} Horizontal linecuts show a series of potential wells, each more than 2~meV deep. 
\textbf{h} \red{Vertical linecuts show a shallow dip and peak behavior, in contrast to \textbf{d}, illustrating the non-trivial interactions between gate layers.}
\textbf{i} Same as \textbf{e}, but now including side-reservoir gates. 
\textbf{j-l} As expected, the reservoir gates have a subtle effect on the inner portion of the device.
\textbf{m} The full gate geometry, also shown Fig.~5b, including the tunnel-barrier gates (yellow).
Here, the Al$_2$O$_3$ layer covering the screening gates is shown (gray), but the Al$_2$O$_3$ layer covering the plunger gates is not shown, for clarity.
\textbf{n-p} The addition of the barrier gates suppresses $\Delta E_c$ oscillations in \textbf{o}, as seen by comparing to \textbf{k}. 
This illustrates the effectiveness of gate-layer stacking for reducing strain fluctuations.
} 
\label{FIG_Quad_Grid}
\vspace{-1mm}
\end{figure*}

\begin{figure}
\begin{center}
\includegraphics[width=3.5in]{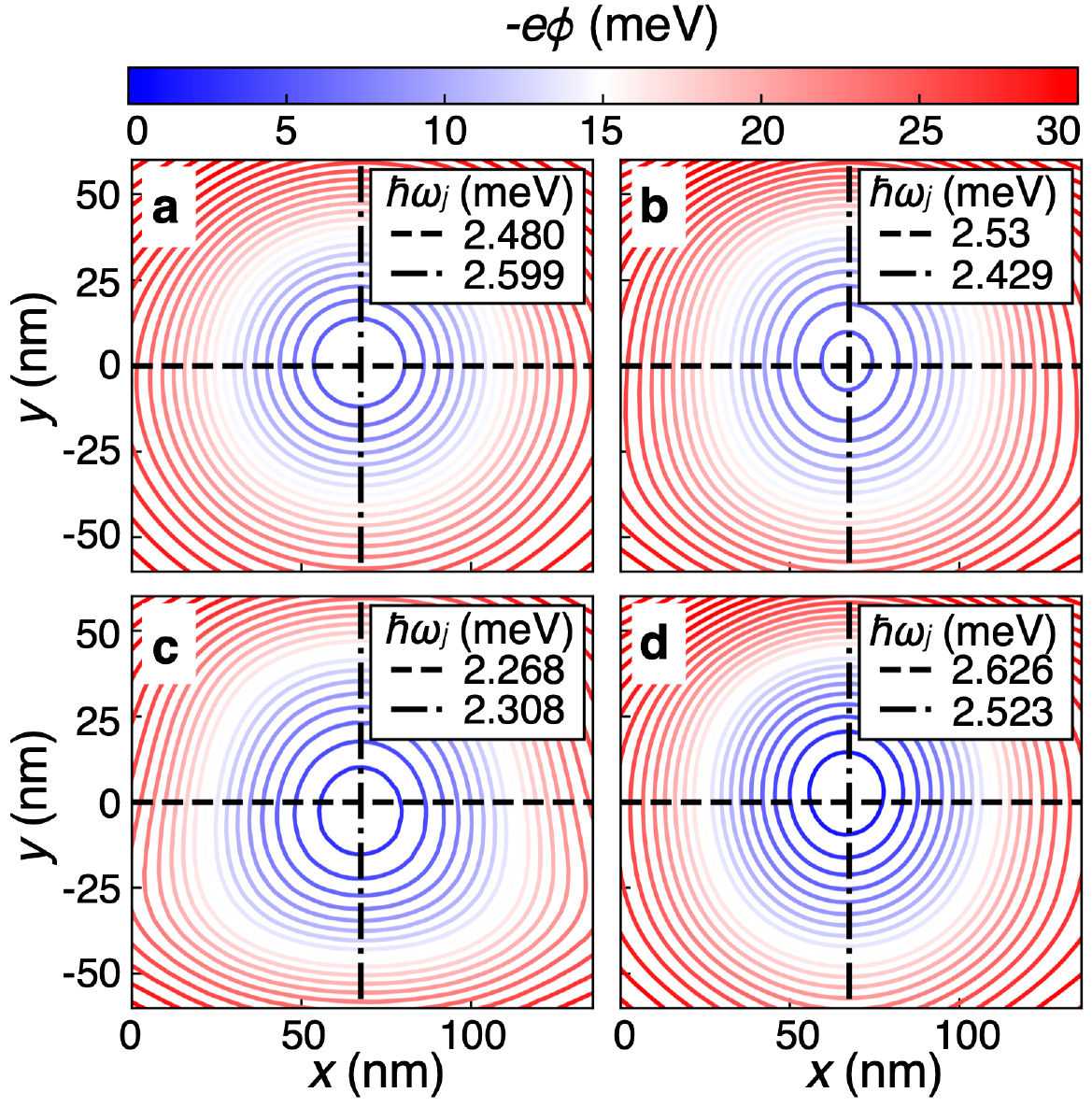}
\end{center}
\vspace{-0.5cm}
\hspace{-1cm}
\caption{
\textbf{Electrostatic and strain contributions to the potential-energy landscape of one of the inner dots of the device shown in Fig.~5.} 
\textbf{a} Simulation results are shown for the electrostatic potential energy -$e\phi$, in the absence of strain, calculated in the Thomas-Fermi approximation~\cite{Stopa} using COMSOL Multiphysics \cite{Comsol}.
Here, gate voltages have been adjusted to achieve single-electron filling of the simulated dot, while depleting the other dots, and filling the nearby charge reservoirs to a density of $2.8\times 10^{11}$~cm$^{-2}$. 
The resulting electrostatic profile is nearly circular, with confinement energies determined by fitting to a parabolic potential: $\{\hbar \omega_x, \hbar \omega_y\} = \{2.480,2.599\}$~meV. 
\textbf{b-d} Same as \textbf{a}, except we now include strain-induced $\Delta E_c$ variations.
Three cases are considered: \textbf{b} Al gates with 60 MPa of depositional stress, \textbf{c} Pd gates with -890 MPa of depositional stress, and \textbf{d} Pd gates with 560 MPa of depositional stress. 
In each case, the electrostatic contribution to the confinement potential dominates over the strain contribution, as evidenced by the small change in confinement energies compared to \textbf{a}, implying that the strain does not strongly affect dot formation. 
In all three cases, however, strain breaks the circular symmetry (reducing $\hbar\omega_y$ with respect to $\hbar\omega_x$), causing the dot to elongate along $\hat y$.
\red{Additionally, for dots with smaller electrostatic confinement energies (e.g., $\hbar \omega_{x,y} \approx 1$~meV), the strain and electrostatic potentials are comparable in magnitude.}
}
\label{FIG_Dot_Confinements}
\vspace{-1mm}
\end{figure}

\begin{figure}
\begin{center}
\includegraphics[width=4in]{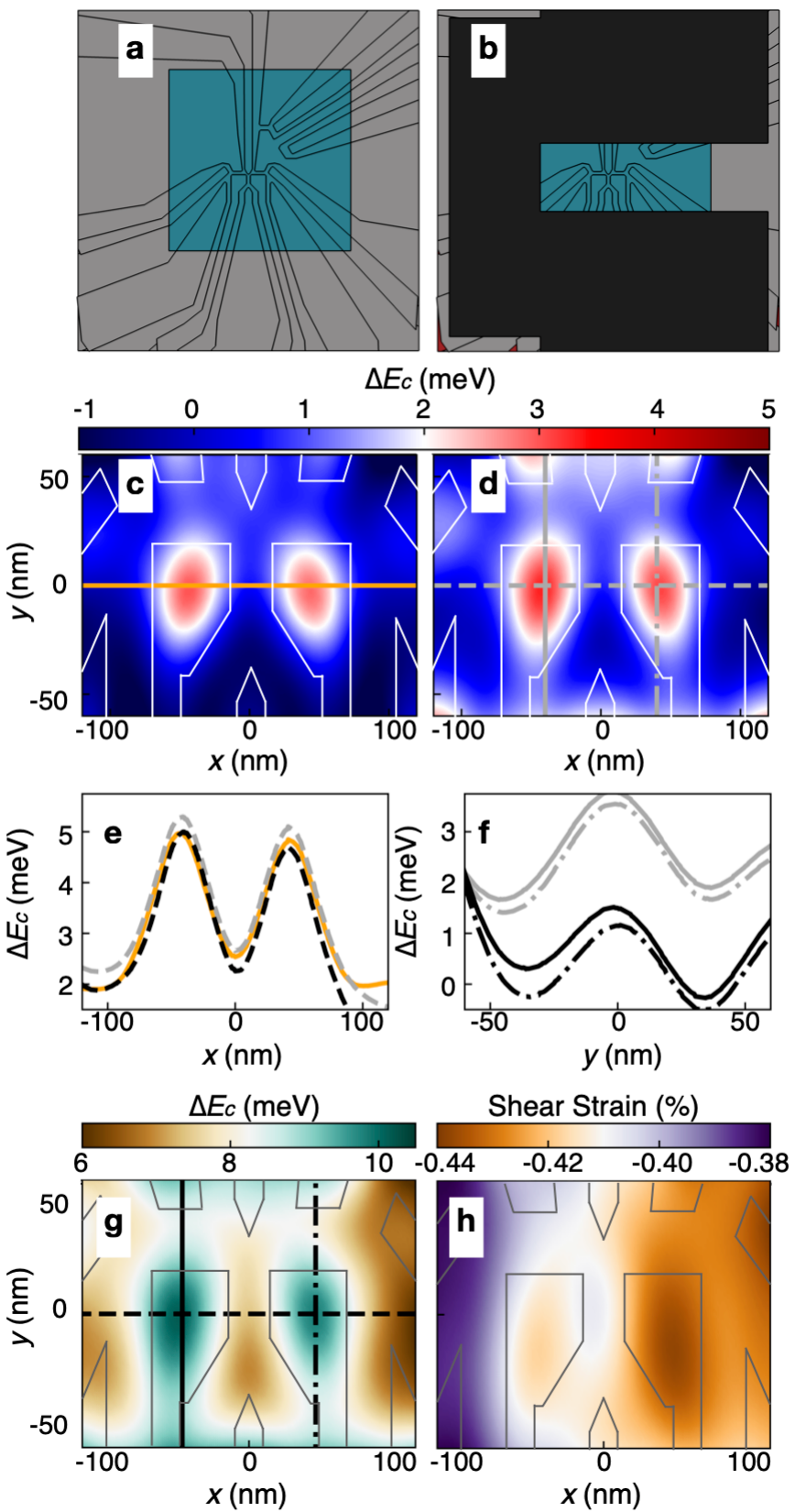}
\end{center}
\vspace{-0.5cm}
\hspace{-1cm}
\caption{
\textbf{Additional simulations of the device shown in Fig.~4.}
\textbf{a} The same gate geometry as in Fig.~4\textbf{b}.
\textbf{b} Again the same device, but now including a patterned cobalt micromagnet of thickness 200~nm. 
\red{Two values for the depositional stress of Co were considered in our simulations, reflecting two different fabrication methods: 150~MPa was used for a sputtering method \cite{Pletea2006,Graciano2019}, and 1300 MPa was used for an evaporation method \cite{Lieven}.}
The temperature-adjusted CTE value for Co, assuming a temperature drop from 293.15~K to 173~K, is given by \SI{12.14e-6}{\per\kelvin}~\cite{Barron1980}.
Although we do not know the CTE corrections for lower temperatures, the trend in~\cite{Barron1980} suggests that this CTE value should give an upper bound on contraction effects.
\textbf{c,d} Results for strain-induced variations of $\Delta E_c$, for the devices shown in \textbf{a} and \textbf{b}, respectively. The device in \textbf{d} has a depositional stress of 150 MPa. 
\textbf{e,f} Linecuts of results shown in \textbf{c}, \textbf{d}, and \textbf{g}:
\textbf{e} horizontal linecuts; \textbf{f} vertical linecuts, with matching color and line styles. 
\red{\textbf{g} Result for strain-induced variation of $\Delta E_c$, for the device shown in \textbf{b}, with a depositional stress of 1300~MPa assumed for the micromagnet. 
\textbf{h} Shear strain produced by the micromagnet, for the device shown in \textbf{b}, with depositional stress of 1300~MPa. }
}
\label{FIG_SingLayer_Appendix}
\vspace{-1mm}
\end{figure}

\begin{figure}
\begin{center}
\includegraphics[width=7in]{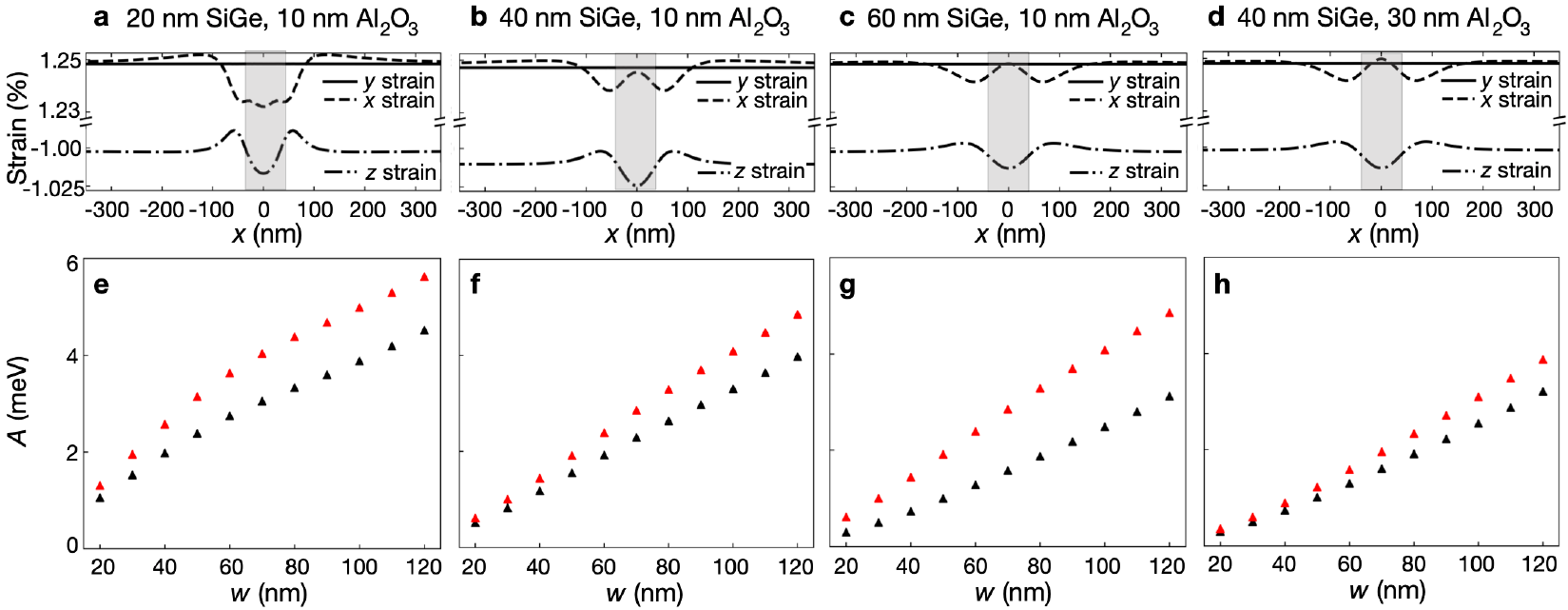}
\end{center}
\vspace{-0.5cm}
\hspace{-1cm}
\caption{
\textbf{Dependence of strain fluctuations on spacer-layer thickness.}
Here, we perform parameter sweeps for the device shown in Fig.~2 of the main text, while also considering different thicknesses of the materials layers below the gate layer.
In Fig.~2, we considered only the case of a 40~nm SiGe spacer layer, with a 10~nm Al$_2$O$_3$ layer on top of it, below the gate layer.
Here, we also consider the cases of a~20 nm SiGe spacer and a 60~nm SiGe spacer, as indicated.
Additionally, we consider a 40~nm SiGe spacer (like in Fig.~2), with a thicker 30~nm Al$_2$O$_3$ layer.
\textbf{a}-\textbf{d} The diagonal strain components $\varepsilon_{xx}$, $\varepsilon_{yy}$, and $\varepsilon_{zz}$ are computed in the plane of the 2DEG and plotted analogously to Fig.~2c, for an Al wire of width 80~nm and height 60~nm, covered on the top and sides by a thin 2~nm Al$_2$O$_3$ layer. 
In these different panels, only the heterostructure is varied, while the gate shape remains constant.
In \textbf{c} and \textbf{d}, note that the total spacer thickness (SiGe spacer plus oxide) is the same (70~nm), and that consequently, no qualitative differences are observed in their strain behavior. 
Generally, we find that a thicker spacer layer yields a lower $z$ strain below the gate, as expected. 
(Far away from the gate, the strain is determined by lattice mismatch in the quantum well.)
\textbf{e}-\textbf{h} 
Here we perform parameter sweeps, varying only the gate width $w$; we compute the amplitude $A$ of the energy fluctuations the same way as in Fig.~2.
The heterostructures are the same as those indicated directly above, in \textbf{a}-\textbf{d}.
We report the results using the same color and marker scheme as Fig.~2d, where the red triangles correspond to Al wires with 60~MPa of depositional stress and black triangles correspond to Al wires with no depositional stress.
Here, we find that a thicker spacer layer generally suppresses the amplitude $A$ of the energy fluctuations.}
\label{FIG_SingWire_Appendix}
\vspace{-1mm}
\end{figure}

\begin{figure}
\begin{center}
\includegraphics[width=7in]{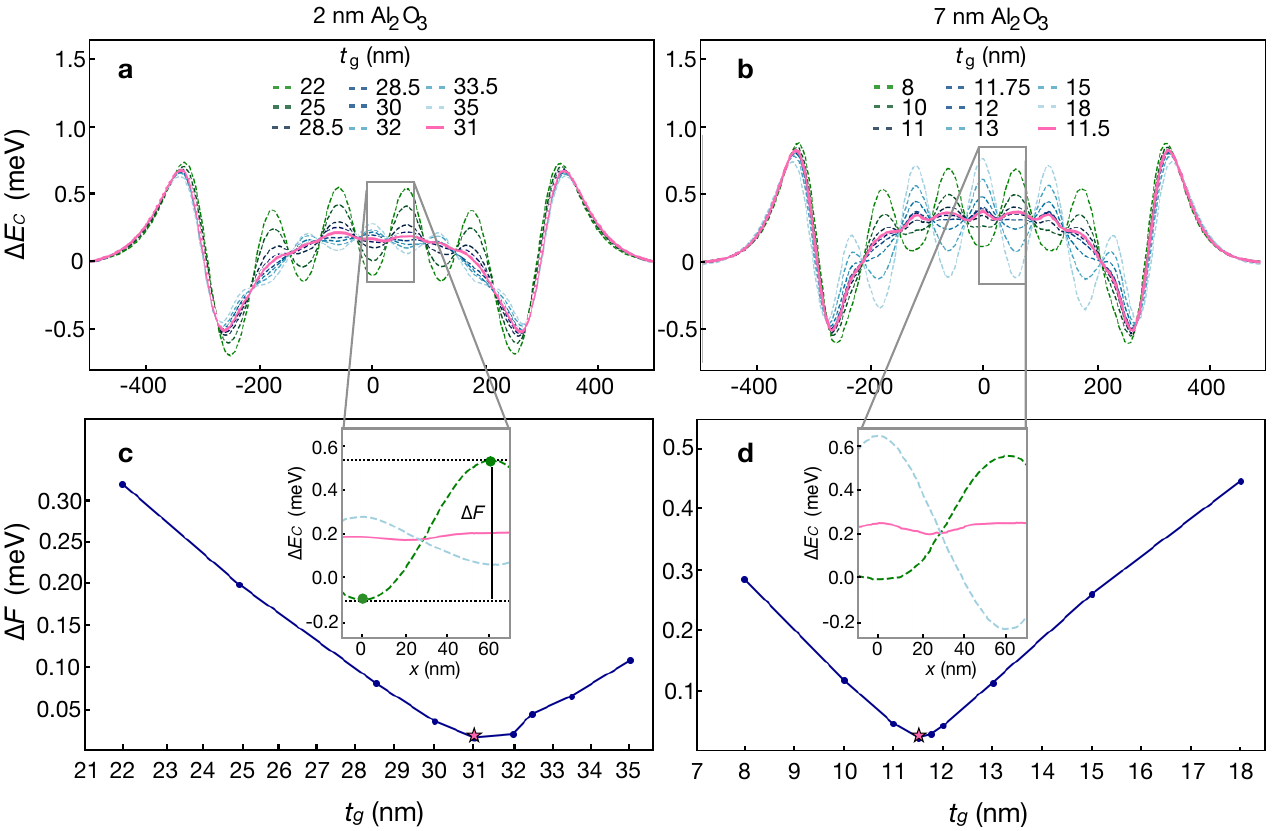}
\end{center}
\vspace{-0.5cm}
\hspace{-1cm}
\caption{
\textbf{Robustness of the suppressed strain fluctuations}
In Fig.~3 of the main text, we showed that certain combinations of gate and oxide thicknesses could strongly suppress the short-range strain fluctuations below the gates.
Here, we explore the robustness of these results.
\textbf{a}, \textbf{b} We consider the oxide thicknesses $t_\text{ox}=2$ and 7~nm, performing simulations of the potential energy fluctuations, similar to Figs.~3c and d.
Here, we simulate a range of $t_g$ values near the optimal values, given by $t_g=31$~nm in \textbf{a} and $t_g=11.5$~nm in \textbf{b}.
\textbf{c}, \textbf{d} We report the magnitude of the fluctuations $\Delta F$ in \textbf{a} and \textbf{b}, as defined in the insets. 
The range of the analysis method shown here is $-10 \leq x \leq 70$~nm.
The stars in \textbf{c} and \textbf{d} correspond to the optimal values of $t_g$, at which $\Delta F$ is minimized.
These results show that fluctuations are suppressed over a robust $t_g$ range of several nanometers.
The details of the results depend on $t_g$, $t_\text{ox}$, and the details of the gate geometry.
Hence, simulations are needed to determine the device parameters that minimize $\Delta F$ for a given geometry.}
\label{FIG_MultiWireProcess_Appendix}
\vspace{-1mm}
\end{figure}
  
\clearpage

\begin{figure}
\begin{center}
\includegraphics[width=6in]{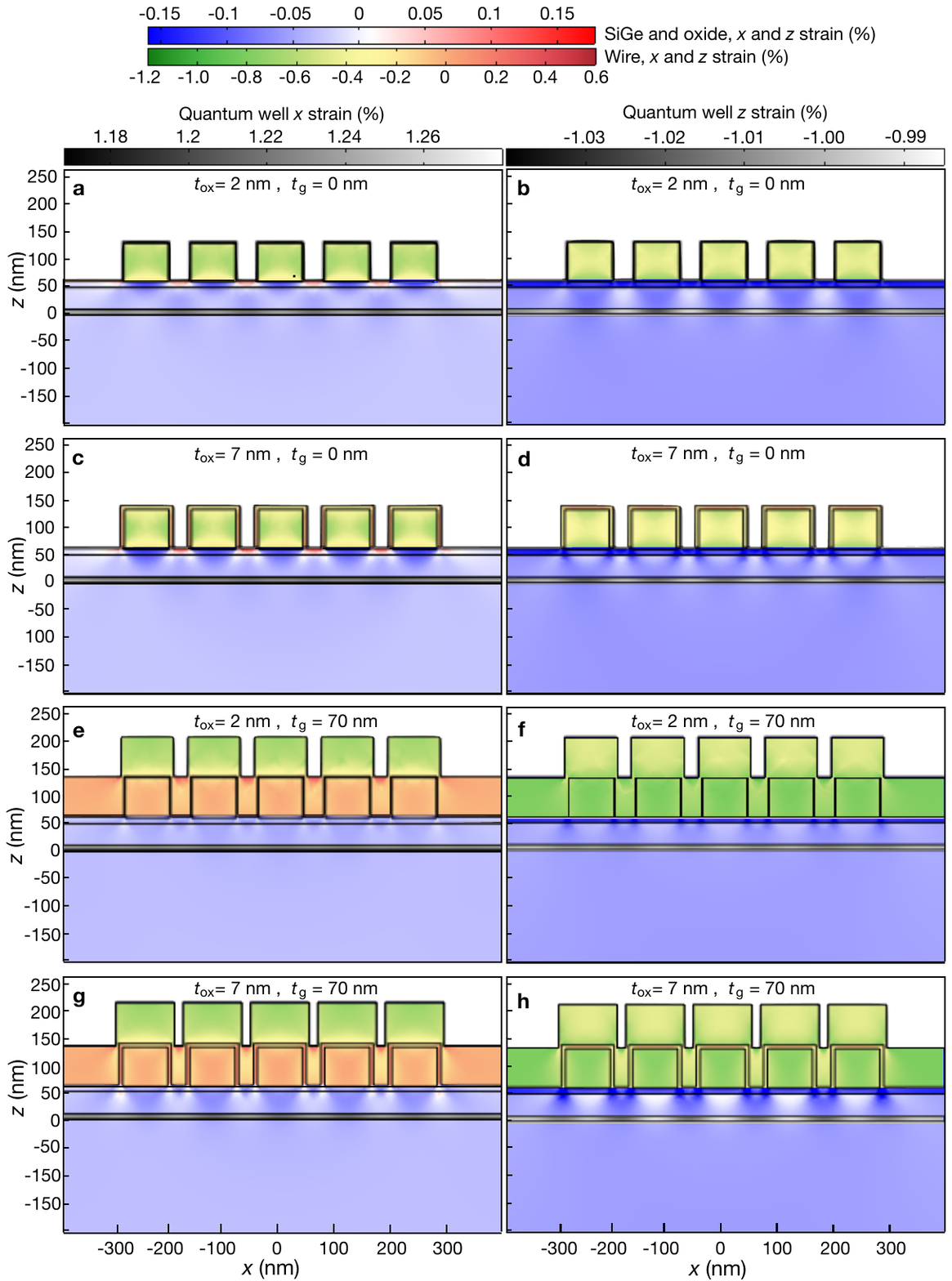}
\end{center}
\vspace{-0.5cm}
\hspace{-1cm}
\caption{
\textbf{Cross-sectional 2D depiction of the strain fields responsible for the energy fluctuations shown in Fig.~3 of the main text.}
Due to translational symmetry along $\hat y$, we only show cross-sectional views in the $x$-$z$ plane.
Here, the first column shows results for the $x$ strain ($\varepsilon_{xx}$), while the second column shows results for the $z$ strain ($\varepsilon_{zz}$); note that these strains are depicted using different color scales, as indicated above.
Focusing on the $z$ strain, which mainly determines the potential energy fluctuations, we observe strain variations consistent with our understanding of gate- vs oxide-driven behavior.
According to our previous analysis, we expect the cases with $t_g=0$ to fall into the gate-driven regime, while $t_g=70$~nm should fall into the oxide-driven regime.
To confirm this behavior in the $z$ strain, we note the color changes in the quantum well, directly below the five wires: here we observe negative strain fluctuations (dark shading) for the $t_g=0$ cases, but positive strain fluctuations (light shading) for the $t_g=70$~nm cases, where the latter is more pronounced for the thicker oxide ($t_\text{ox}=7$~nm).}
\label{FIG_MultiWireCrossSection_Appendix}
\vspace{-1mm}
\end{figure}

\clearpage

\begin{figure}
\begin{center}
\includegraphics[width=6in]{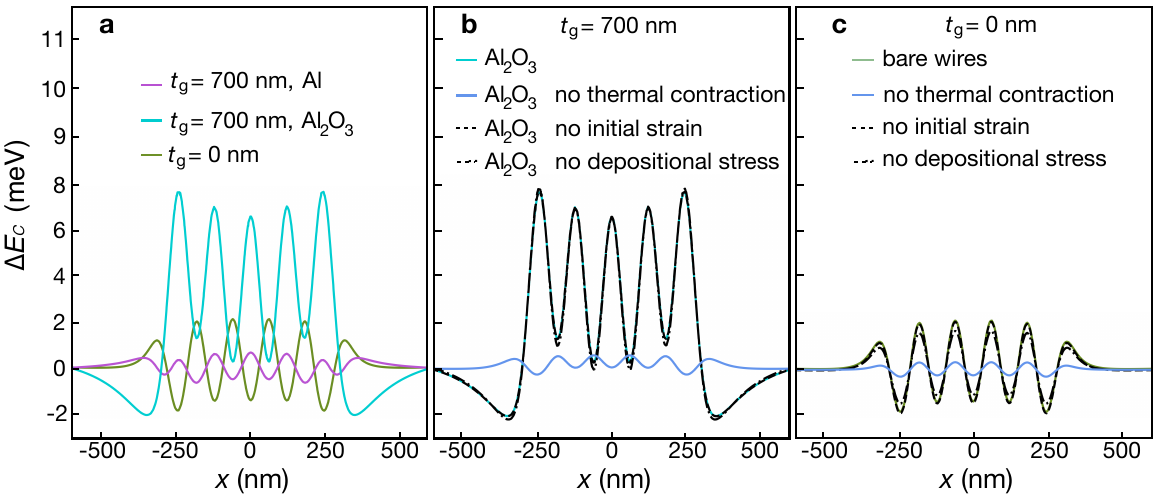}
\end{center}
\vspace{-0.5cm}
\hspace{-1cm}
\caption{
\red{\textbf{Exploration of the dominant sources of strain.}
Additional simulations are now performed for the same geometry as  Fig.~3 of the main text; however, here, different materials are used for the global gate, including Al and Al$_2$O$_3$, and we also consider the case of no global gate ($t_g=0$). 
\textbf{a} Here, the sign of the $\Delta E_c$ oscillations for both the thick Al global gate (purple curve) and the thick Al$_2$O$_3$ gate (cyan curve) are the same, indicating that both devices are in the oxide-driven regime. The configuration with $t_\text{g} = 0$~nm falls into the gate-driven regime, as expected. 
\textbf{b} Here, we investigate the separate contributions from each of the three main strain sources considered in this work: thermal contraction, lattice-matching strain, and depositional stress.
The device considered in this case is the same as in \textbf{a}, and we now consider only the case of a thick Al$_2$O$_3$ top gate, with $t_\text{g} = 700$~nm. We probe the different strain sources by starting with a complete device simulation (cyan curve) and then turning off each of the strain sources. 
(Note again that thermal contractions were considered -- or turned off -- for both the Al and Al$_2$O$_3$ layers of the device. However, depositional stress was considered only for the Al material. See the main text for discussion.)
For this geometry, it is clear that thermal contraction is the dominant effect.
However, in cases where depositional stress is very large (for example, for the Co micromagnet in Fig.~S6), the depositional stress may dominate. \textbf{c} Here, a similar strain-source comparison is performed for a device with no global top gate ($t_\text{g} = 0$~nm). Interestingly, the oscillations observed in several of the curves have an opposite sign as the oscillations observed in \textbf{b}.  }
}
\label{FIG_MultiWireContinued}
\vspace{-1mm}
\end{figure}

\end{document}